\documentclass[hidelinks,preprint,3p,times,twocolumn]{elsarticle}

\usepackage{amssymb}
\usepackage{amsmath}

\usepackage{graphicx}

\journal{NIM-A}

\usepackage{orcidlink}
\usepackage{lmodern}
\usepackage{graphicx}
\usepackage{hyperref}

\usepackage[table,xcdraw]{xcolor}

\usepackage{nolbreaks}

\usepackage[separate-uncertainty,exponent-product=\cdot,range-units=single,range-phrase=--]{siunitx}
\DeclareSIUnit{\neqpcm}{\si{n_{eq}\per\centi\meter\squared}}
\DeclareSIUnit{\npcm}{\si{n\per\centi\meter\squared}}
\DeclareSIUnit{\nneq}{\si{n_{eq}}}
\DeclareSIUnit{\ppcm}{\si{p\per\centi\meter\squared}}
\DeclareSIUnit{\epcm}{\si{e^-\per\centi\meter\squared}}
\DeclareSIUnit{\electron}{e^-}
\DeclareSIUnit{\micron}{\micro\meter}
\DeclareSIUnit{\microns}{\micro\meter}
\DeclareSIUnit{\ohmcm}{\ohm\centi\meter}
\DeclareSIUnit{\clight}{\text{\ensuremath{c}}}
\DeclareSIUnit\barn{b}
\DeclareSIUnit{\centimeter}{\centi\meter}

\usepackage{xspace}

\newcommand{\IV}{$I$--$V$\xspace}
\newcommand{\CV}{$C$--$V$\xspace}
\newcommand{\BiOi}{$\text{B}_\text{i}\text{O}_\text{i}$\xspace}

\newcommand{\Si}{$\text{Si}_\text{i}$\xspace}
\newcommand{\Bi}{$\text{B}_\text{i}$\xspace}
\newcommand{\Bs}{$\text{B}_\text{s}$\xspace}

\newcommand{\Oi}{$\text{O}_\text{i}$\xspace}

\newcommand{\Neff}{$N_\text{eff}$\xspace}

\newcommand{\NT}{$N_\text{T}$\xspace}

\newcommand{\Tw}{$T_\text{W}$\xspace}

\newcommand{\Vdep}{$V_\text{dep}$\xspace}

\newcommand{\ptype}{$p$-type\xspace}

\newcommand{\electron}{$\text{e}^-$\xspace}

\newcommand{\EA}{$E_\text{A}$\xspace}
\newcommand{\SiiBs}{$\text{Si}_\text{i}\text{B}_\text{s}$\xspace}

\newcommand{\figuresref}[1]{\hyperref[#1]{\textcolor{black}{Figures~\ref{#1}}}}
\newcommand{\tablesref}[1]{\hyperref[#1]{\textcolor{black}{Tables~\ref{#1}}}}
\newcommand{\equationsref}[1]{\hyperref[#1]{\textcolor{black}{Equations~\ref{#1}}}}

\usepackage{makecell}
\usepackage{tabularx}
\usepackage{caption}

\usepackage{placeins}
\usepackage[T1]{fontenc}
\usepackage{microtype}

\begin{document}

\begin{frontmatter}



\title{Gain-Layer Project}

\author[a,b]{Niels G. Sorgenfrei\orcidlink{0000-0002-5729-6004}\corref{1}}
\cortext[1]{Corresponding authors}\ead{niels.sorgenfrei@cern.ch}

\author[e]{Anna Rita Altamura}
\author[c]{Cristina Besleaga\orcidlink{0000-0003-3311-869X}}
\author[c]{Georgia Andra Boni\orcidlink{0000-0002-8355-6826}}
\author[f]{Tomas Ceponis\orcidlink{0000-0001-5448-6390}}
\author[d]{Paul Erberk}
\author[d]{Eckhart Fretwurst}
\author[a]{Yana Gurimskaya\orcidlink{0000-0002-2549-4153}}
\author[g]{Kevin Lauer\orcidlink{0000-0002-7474-9239}}
\author[e]{Ludovico Massaccesi\orcidlink{0000-0003-1762-4699}}
\author[a]{Luca Menzio\orcidlink{0000-0002-9697-5608}}
\author[a]{Michael Moll\orcidlink{0000-0001-7013-9751}}
\author[a]{Marie Mühlnikel\orcidlink{0009-0007-6437-3542}}
\author[c]{Andrei Nitescu\orcidlink{0000-0002-4465-2911}}
\author[b]{Ulrich Parzefall}
\author[c]{Roxana-Elena Patru}
\author[f]{Jevgenij Pavlov\orcidlink{0009-0009-8588-1268}}
\author[c]{Ioana Pintilie\orcidlink{0000-0002-3857-8524}\corref{1}}\ead{ioana@infim.ro}
\author[g]{Stephanie Reiss\orcidlink{0000-0003-3224-9102}}
\author[d]{Joern Schwandt\orcidlink{0000-0002-0052-597X}}
\author[e]{Valentina Sola\orcidlink{0000-0001-6288-951X}}

\affiliation[a]{organization={CERN, European Organization for Nuclear Research},
            addressline={Esplanade des Particules 1}, 
            city={Geneva},
            postcode={1211}, 
            country={Switzerland}}
\affiliation[b]{organization={Institute of Physics, Albert-Ludwigs-Universitaet Freiburg},
            addressline={Hermann-Herder-Strasse 3}, 
            city={Freiburg im Breisgau},
            postcode={79104}, 
            country={Germany}}

\affiliation[c]{organization={National Institute of Materials Physics},
            addressline={Atomistilor 405A}, 
            city={Magurele},
            postcode={077125}, 
            country={Romania}}
\affiliation[d]{organization={Institute for Experimental Physics, University of Hamburg},
            addressline={Luruper Chaussee 149}, 
            city={Hamburg},
            postcode={22761}, 
            country={Germany}}
\affiliation[e]{organization={Torino University and National Institute for Nuclear Physics (INFN)},
            addressline={Via Pietro Giuria 1}, 
            city={Torino},
            postcode={10125}, 
            country={Italy}}
\affiliation[f]{organization={Institute of Photonics and Nanotechnology, Vilnius University},
            addressline={Saulėtekio av. 9}, 
            city={Vilnius},
            postcode={10222}, 
            country={Lithuania}}
\affiliation[g]{organization={CiS Forschungsinstitut für Mikrosensorik GmbH},
            addressline={Konrad-Zuse-Str. 14}, 
            city={Erfurt},
            postcode={99099}, 
            country={Germany}}

\begin{abstract}

Gain-layer degradation from exposure to radiation limits the use of Low-Gain Avalanche Diodes (LGADs) in high energy particle physics detector experiments.
Proper understanding of how the gain-layer is altered is not available on a defect level.
Only measurements for materials with much lower effective doping concentrations are available.
The direct study of the gain-layer is not possible with typical defect spectroscopy measurements like Thermally Stimulated Currents (TSC) and Deep-Level Transient Spectroscopy (DLTS).
To combat this problem and gain a better understanding of the processes which degrade LGADs, the Gain-Layer Project was started.

Within this project \num{19050} diodes were produced with various Boron, Phosphorus, Oxygen and Carbon concentrations.
The material used is low-resistivity \ptype Silicon.
The effective doping concentrations are in the order of typical LGAD gain-layers.
These diodes will serve the defect community in the coming years for various studies.

This article introduces this project with detailed descriptions of the diodes, their flavours and their processing, and reports on results from \IV, \CV, SIMS and DLTS measurements on unirradiated diodes.

\end{abstract}

\begin{graphicalabstract}
\includegraphics[width=\textwidth]{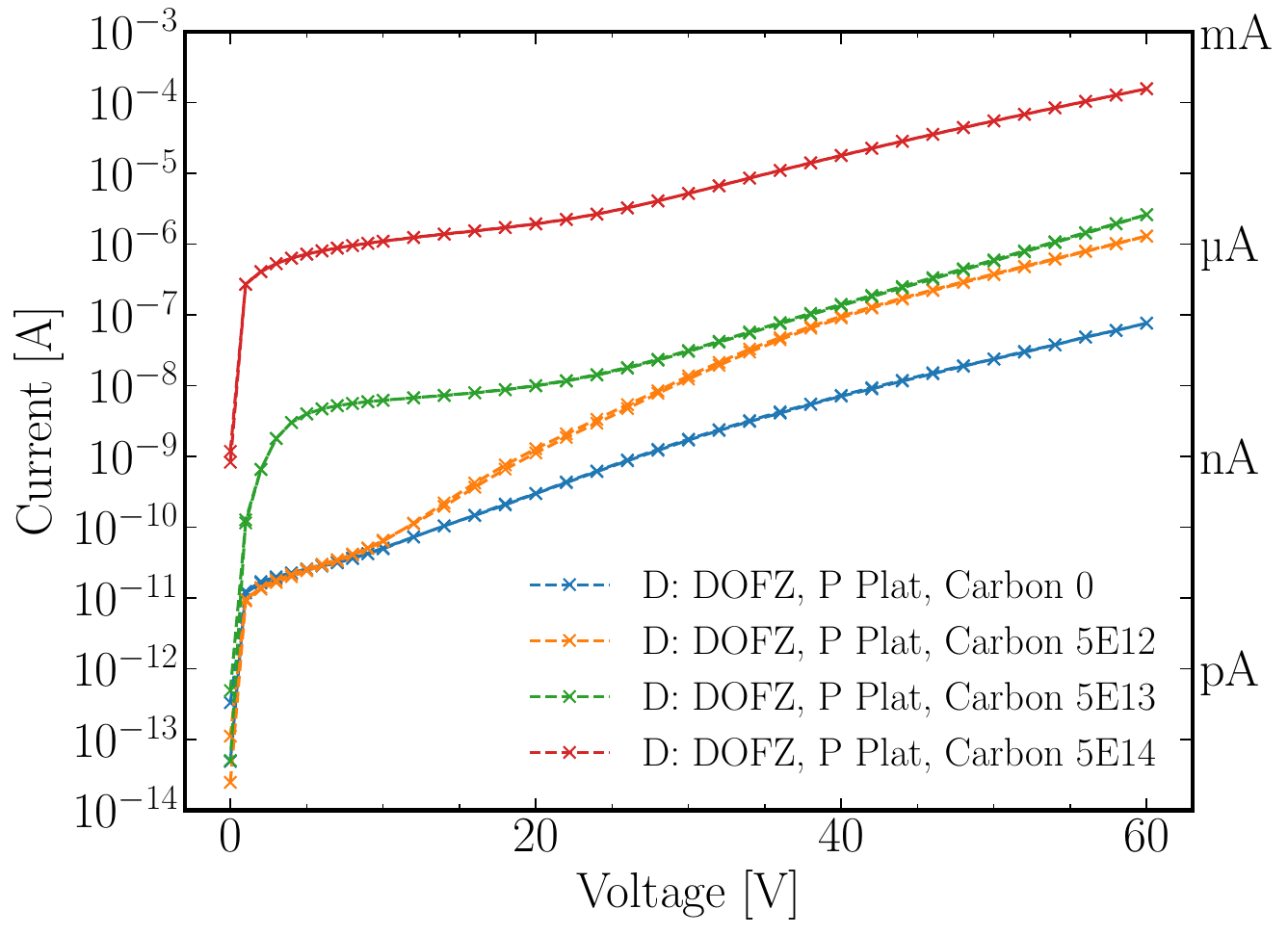}
This figure shows the increase of the leakage currents with higher Carbon implantation doses in diodes on Diffusion-Oxygenated Float-zone \ptype Silicon with Phosphorus co-doping.
\end{graphicalabstract}

\begin{highlights}
\item Research highlight 1\\
Large production of Gain-Layer Project Diodes (\num{19050}) was successful
\item Research highlight 2\\
Unirradiated diodes work well in \IV, \CV and DLTS measurements\\
\item Research highlight 3\\
Most SIMS measurements showed a good agreement with the simulation of the processing
\item Research highlight 4\\
Carbon implantation leads to higher leakage currents, which increase with higher Carbon doses
\item Research highlight 5\\
These diodes can be used for defect studies to understand gain-layer degradation of LGADs with irradiation
\end{highlights}

\begin{keyword}
DLTS \sep defect spectroscopy \sep $p$-type Silicon \sep LGAD \sep GLPD (Gain-Layer Project Diode) \sep electrical characterisation \sep SIMS




\end{keyword}

\end{frontmatter}




\section{Introduction}

Low Gain Avalanche Diodes (LGADs) are sensors with intrinsic gain developed for increased timing requirements in High Energy Physics (HEP) experiments.
With a timing resolution of \SIrange{20}{30}{\pico\second} and an intrinsic gain in the order of 10, LGADs will be used as timing detectors at the upgraded High-Luminosity Large Hadron Collider (HL-LHC) \cite{sadronzinski,Kramberger,ferrero}.
The \mbox{$n^{++}-p^+-p-p^{++}$} structure of LGADs with a thin (\SIrange{1}{2}{\microns}), highly doped $p^+$ region (the gain-layer), provides strong enough electric fields near the junction contact to produce charge multiplication via impact ionisation~\cite{pellegrini}.

The limiting factor for the use of LGADs in harsh radiation environments is the degradation of their gain-layer \cite{Cartiglia,Lange,Menzio}.
A reduction of the gain due to the exposure to radiation lowers the signal-to-noise ratio and degrades the time resolution.
At some point, the gain-layer doping is fully compensated by the formation of radiation-induced defects \cite{ferrero}.
This apparent deactivation of the gain-layer dopants is called Acceptor Removal Effect (ARE), and is directly observed as a variation in the effective doping concentration \Neff.
The Boron dopants (acceptors) are deactivated or compensated by donor-like defects introduced into the Silicon lattice by radiation.

The current understanding of the acceptor removal process in \ptype Silicon is that a Boron Containing Donor (BCD) defect, with an energy level detected at $\sim\SI{0.25}{\electronvolt}$ from the conduction band of Silicon, is formed during irradiation via a reaction between substitutional Boron (\Bs) and interstitial Silicon (\Si).
Two chemical structures are proposed in literature for the identity of the BCD complex.
The Boron-interstitial-Oxygen-interstitial (\BiOi) complex \mbox{\cite{niels_xdefect,anja2, mooney, makarenko2014,ioana7,liao1,liao2}} is a defect that is formed in two steps.
\Bs atoms first switch places with Silicon interstitials becoming Boron interstitials (\Bi) which then migrate in the crystal and combine with the abundant Oxygen interstitials (\Oi) present in Silicon forming the \BiOi defect.
The Silicon-interstitial-Boron-substitutional (\SiiBs) complex \mbox{\cite{lauer,flototto,ioana9}} accounts for a direct reaction between the negatively charged acceptor dopant and the positively charged \Si atom created by irradiation. 
Both \BiOi and \SiiBs defects have donor energy levels in the bandgap of Silicon and thus, both can act as BCD defects being responsible for the ARE. 
Presently, there are no conclusive experiments to clarify the chemical structure of the detected BCD center.
An effective way to mitigate the gain loss is Carbon co-implantation \cite{ferrero}, but the improvement in radiation hardness is still not enough to cover the lifetime of future HEP experiments, where extreme radiation fluences are expected.

To understand the gain-layer degradation, a direct defect spectroscopy measurement of the gain-layer is needed.
However, Deep-Level Transient Spectroscopy (DLTS), the most powerful technique for characterising electrically active defects, is unable to probe the complex structures of LGADs \cite{LGAD_TSC}. 
The Thermally Stimulated Currents (TSC) method can probe the LGAD structures \cite{ioana9}, but for a quantitative evaluation of the radiation damage in the gain-layer only, the doping profile in the implanted $p^+$ thin layer, the  temperature dependence of the gain  as well as the damage in the bulk of the measured LGADs has to be well determined.  
Consequently, most studies have so far relied on measurements of standard \mbox{$n^{++}-p$} pad diodes with resistivities ranging from \SI{50}{\ohmcm} to several \si{\kilo\ohmcm}, and irradiated at a maximum \SI{1}{\mega\electronvolt} neutron equivalent fluence $\Phi_\text{eq}$ of \SI{e15}{\per\centi\meter\squared} \mbox{\cite{anja2,makarenko2014,ioana7,liao1,liao2,ioana9,anja1,Moll:2020On}}.

Radiation-induced defects after higher irradiation fluences were not investigated so far. 
Further experimental measurements are required, together with new model development.
New techniques that are able to handle large defect concentrations, such as Fourier Transform Infrared (FTIR) and Photo-luminescence (PL) spectroscopies, should be employed. 
Also, specially-designed samples should be fabricated for these studies.
Moreover, the change of the fundamental semiconductor properties, e.g. the carrier mobilities, lifetimes and impact ionisation, at extreme fluences are not precisely known, although they are needed for any detector design.

The Gain-Layer Project was envisioned to address these issues. 
It aims to study defects in specially-engineered \ptype pad diodes with a highly-doped bulk, which mimics the gain-layer of LGADs. 
We refer to these diodes as Gain-Layer Project Diodes (GLPDs).
The project started in the framework of the RD50/DRD3 collaborations \cite{RD50,drd3}, and produced \num{19050} diodes with bulk doping close to the doping levels used for establishing the gain-layers in LGADs, specifically designed for defect studies.
Thus, planar \mbox{$n^{++}-p^+-p^{++}$} defect engineering Silicon diodes mimicking the gain-layer of LGADs with 3 different Carbon implantation doses, Boron (\SIlist{2;10}{\ohmcm}), Phosphorus (for a partial compensation of Boron doping in the \SI{2}{\ohmcm} wafers) and Oxygen content have been fabricated.
They are processed on standard and Oxygen-Diffused Float-zone Silicon (FZ and DOFZ) wafers of \SIlist{2;10}{\ohmcm} resistivity \ptype substrate.
They are designed to be thoroughly investigated from the microscopic point of view over a large fluence range, using DLTS, TSCap and TSC.

TSC requires fully-depleted samples. 
However, these highly-doped substrates are expected to have very large depletion voltages. 
To allow full-depletion of a controlled volume, diodes with a thin ($\sim$\SI{2}{\micron}) layer with \SI{10}{\ohmcm} resistivity are produced out of \SI{2}{\ohmcm} substrates by compensating the Boron doping of the bulk with a Phosphorus implant at different energies. 
This solution also allows to study the Donor Removal Effect (DRE) and the possibility to exploit it to mitigate the ARE in gain-layers with Boron-Phosphorus co-implantation.

GLPDs will be irradiated at different fluences, ranging from $\sim$\SI{e13}{\neqpcm} to over \SI{e16}{\neqpcm}. 
Different experimental methods will be used for defect characterisation as appropriate: DLTS up to \SI{e15}{\neqpcm}, TSC and TSCap up to \SI{e16}{\neqpcm}, and FTIR and PL spectroscopies above \SI{e16}{\neqpcm}.
Charge Collection Efficiency (CCE), Transient Current Technique (TCT), charge carrier lifetime, Van der Pauw resistivity and Hall mobility measurements will also be employed.

Further on, the parameters determined from experiments will be used as reference values or inputs for modelling both the defect generation and kinetics and the device properties.
The goal of the project is to understand the ARE in irradiated \ptype Silicon, parametrising it for various contents of Boron, Carbon, Phosphorus and Oxygen impurities and irradiation fluences.
This way, we can find solutions to maximise the radiation hardness of LGADs through defect and material engineering, device engineering and optimisation of operational conditions.

This article introduces the project, reports its status and the \IV, \CV, SIMS and DLTS measurement results on as-processed samples.

\section{Gain-Layer Project Diodes}
CiS Forschungsinstitut für Mikrosensorik GmbH~\cite{cis} produced \num{19050} GLPDs distributed on \num{25} wafers.
A total of six distinct flavours were produced.
The diodes are made from \ptype Silicon wafers of two resistivities: \SI{2}{\ohmcm} and \SI{10}{\ohmcm}.
The former are \SI{250}{\micron} thick, while the latter are \SI{525}{\microns} thick. 
Additionally, both FZ and DOFZ wafers were used. 
Finally, some wafers are processed with an additional Phosphorus plateau implant.

\begin{figure*}
    \centering
    \includegraphics[width=\linewidth]{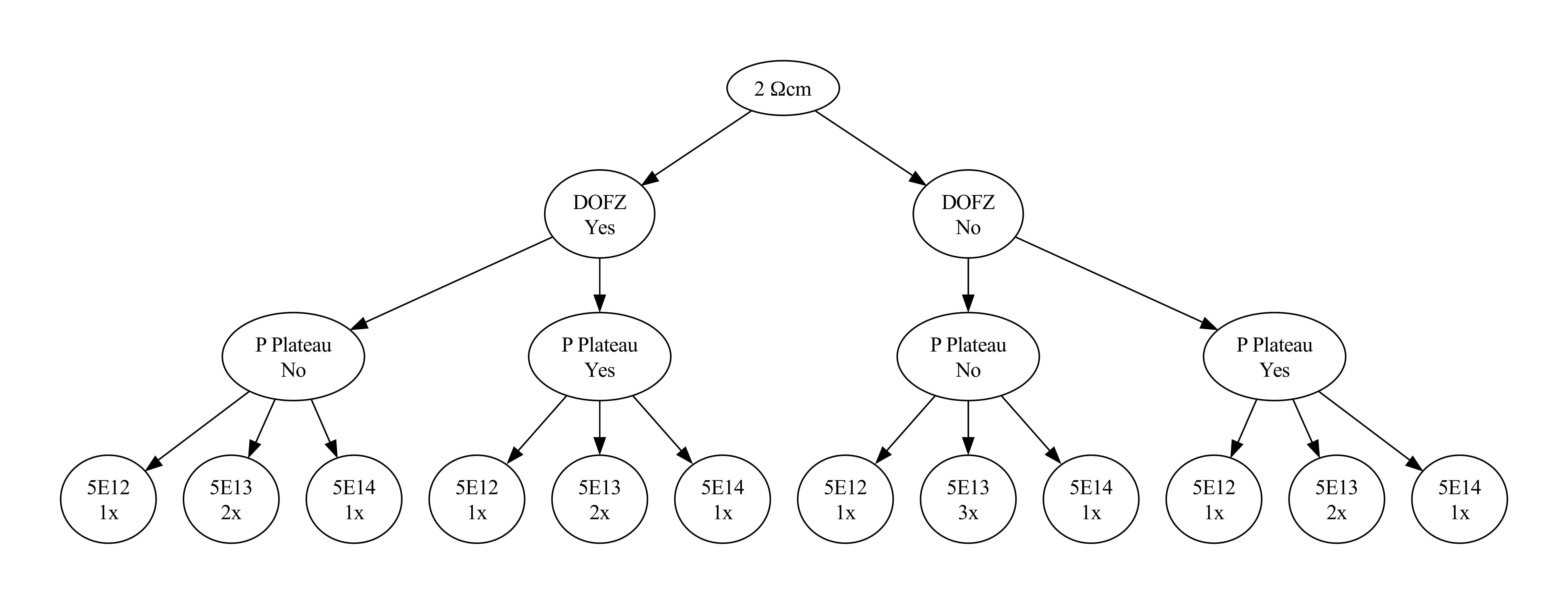}
    \caption{Flavour tree for the \SI{2}{\ohmcm}, \SI{250}{\microns} thick wafers. The last row indicates the Carbon implantation dose in units of \si{\per\centi\meter\squared} and the number of wafers produced for each specific combination.}
    \label{fig:tree_2}
\end{figure*}

\autoref{tab:flavours} gives a summary of the flavours, while \figuresref{fig:tree_2} and \ref{fig:tree_10} show the split of the various parameters.
\autoref{tab:glpd} in the appendix lists all wafers and their specific flavour.

For each flavour, the wafers were implanted with three different Carbon doses: \SIlist{5e12;5e13;5e14}{\per\centi\meter\squared}.
The top right quadrant of each wafer was left without Carbon which was achieved by gluing on a sacrificial Silicon wafer quarter during the Carbon implantation.
For each flavour with the middle Carbon dose, two (for flavour A three) identical wafers were produced.
For the lowest and highest Carbon dose, only one wafer per flavour was produced.

\begin{table}
\centering
\resizebox{\linewidth}{!}{%
\begin{tabular}{ccccc}
\hline
\small{Flavour} & \small{Resistivity} & \small{DOFZ} & \small{Phosphorus} & \small{Carbon}\\
 & [\si{\ohm\centi\meter}] & & \small{Plateau} & [\si{\per\centi\meter\squared}]\\\hline
A & 2 & \color[HTML]{CB0000} no & \color[HTML]{CB0000} no & \footnotesize{0, \num{5E12}, \num{5E13}, \num{5E14}}\\
B & 2 & \color[HTML]{CB0000} no & \color[HTML]{009901} yes & \footnotesize{0, \num{5E12}, \num{5E13}, \num{5E14}}\\
C & 2 & \color[HTML]{009901} yes & \color[HTML]{CB0000} no & \footnotesize{0, \num{5E12}, \num{5E13}, \num{5E14}}\\
D & 2 & \color[HTML]{009901} yes & \color[HTML]{009901} yes & \footnotesize{0, \num{5E12}, \num{5E13}, \num{5E14}}\\
E & 10 & \color[HTML]{CB0000} no & \color[HTML]{CB0000} no & \footnotesize{0, \num{5E12}, \num{5E13}, \num{5E14}}\\
F & 10 & \color[HTML]{009901} yes & \color[HTML]{CB0000} no & \footnotesize{0, \num{5E12}, \num{5E13}, \num{5E14}}\\
\hline
\end{tabular}
}
\caption{List of the six flavours of GLPDs.}
\label{tab:flavours}
\end{table}

\begin{figure}
    \centering
    \includegraphics[width=\linewidth]{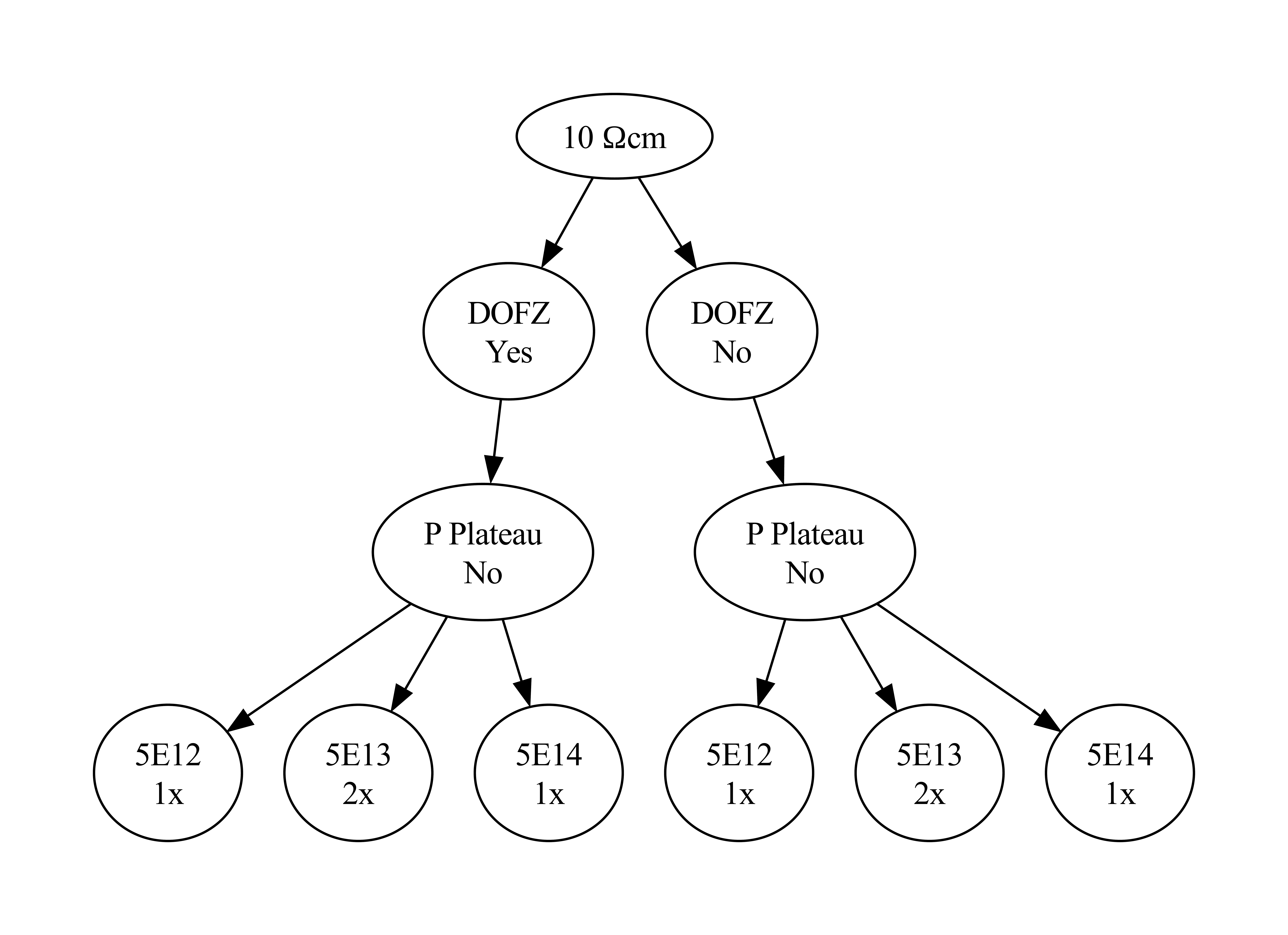}
    \caption{Flavour tree for the \SI{10}{\ohmcm}, \SI{525}{\microns} thick wafers. The last row indicates the Carbon implantation dose in units of \si{\per\centi\meter\squared} and the number of wafers produced for each specific combination.}
    \label{fig:tree_10}
\end{figure}

\begin{figure*}[t]
    \centering
    \includegraphics[width=\linewidth]{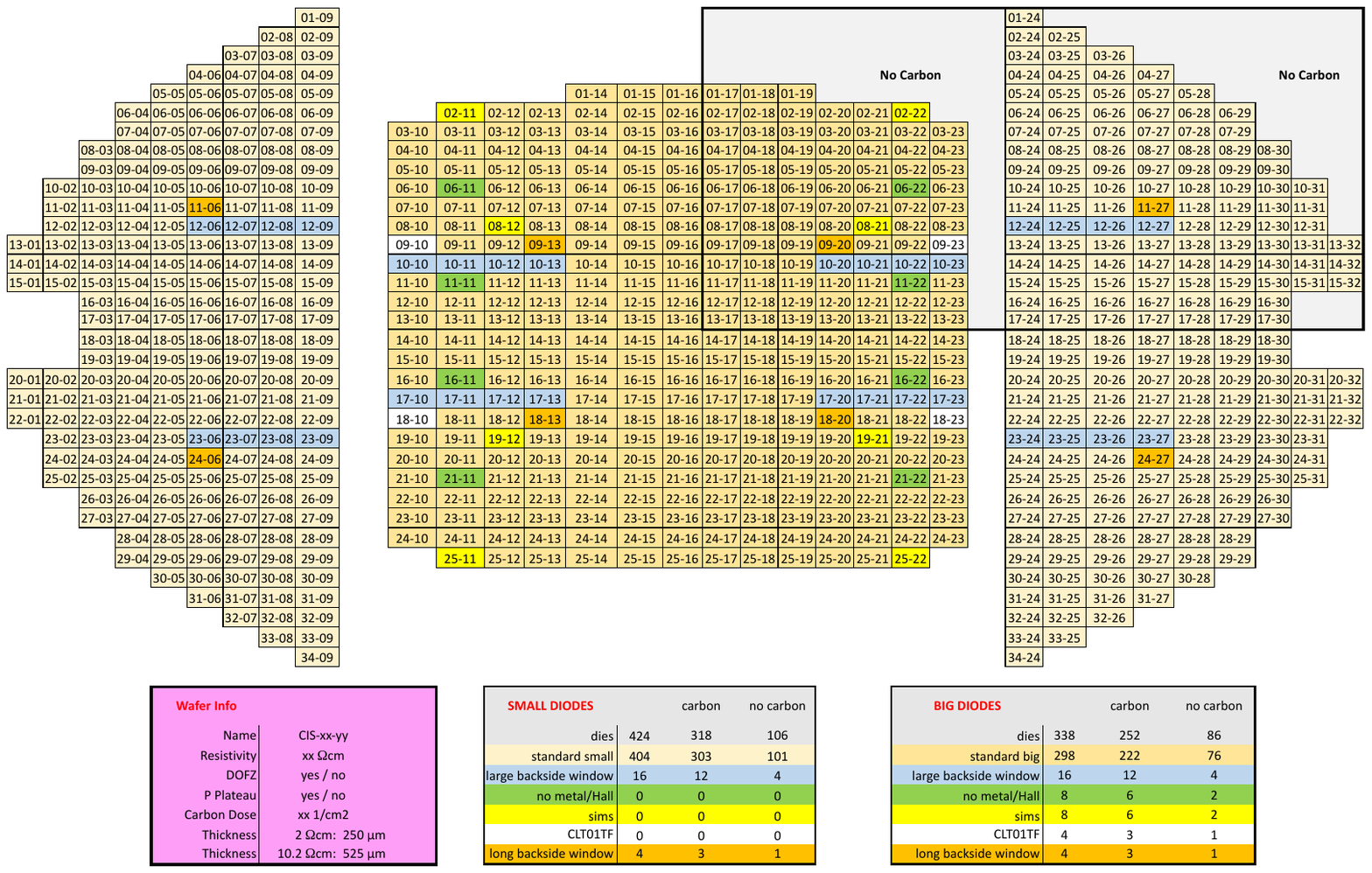}
    \caption{Wafer layout.}
    \label{fig:wafer_layout}
\end{figure*}

\autoref{fig:wafer_layout} shows the schematic layout of the wafers.
The middle part of each wafer contains large diodes with a pad size of \SI{6.25}{\milli\meter\squared} (this is the surface encased by the guard ring). 
The left and right side of each wafer contains small diodes with a pad size of \SI{1.5625}{\milli\meter\squared}.
As can be seen in \autoref{fig:wafer_layout}, some special structures were produced.
The bulk of the wafer is filled with standard pad diodes, while a small fraction has larger openings in the metal, both for the front and backside.
Some structures have no metal layers to allow Hall measurements, and some are dedicated to Secondary-Ion Mass Spectroscopy (SIMS) measurements.


A detailed description of the production process is given in the following.

\section{Simulation}
To tailor the processing parameters, the dopant profiles were simulated using the Atlas software from Silvaco \cite{Silvaco} and SRIM \cite{SRIM,SRIM2}.
There were two goals: creating a Phosphorus plateau in some \SI{2}{\ohmcm} wafers, directly below the $n-p$ junction, to generate a compensated region with a resistivity of \SI{10}{\ohmcm}, and implanting Carbon into the $n-p$ junction.

\begin{figure}
    \centering
    \includegraphics[width=\linewidth]{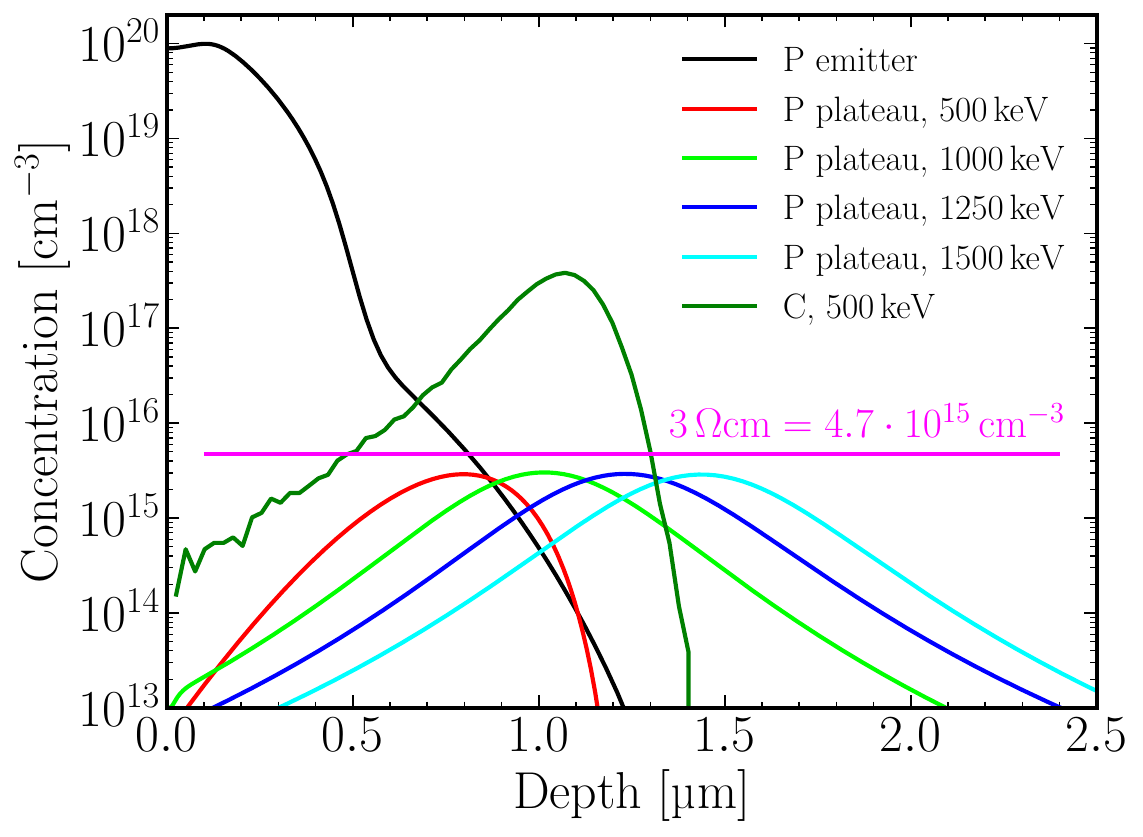}
    \caption{Simulated Phosphorous and Carbon profiles.}
    \label{fig:conc_sim}
\end{figure}

\autoref{fig:conc_sim} shows the results of the simulation.
The simulated implantation energies for Phosphorus are \SIlist{500;1000;1250;1500}{\kilo\electronvolt}. 
At \SI{500}{\kilo\electronvolt}, a Phosphorous dose of \SI{1.2e11}{\per\centi\meter\squared} is used and an annealing step of \SI{20}{\second} at \SI{1000}{\celsius} is simulated. 
For the other implantation energies, a dose of \SI{1.6e11}{\per\centi\meter\squared} is used and an annealing step of \SI{5}{\minute} at \SI{1000}{\celsius} is simulated. 
The Carbon profile given in the figure is simulated using SRIM without an annealing step after the implantation. 
The simulated Carbon implantation dose is \SI{1e13}{\per\centi\meter\squared}.

\section{Processing}
According to the flavour tree in \autoref{fig:tree_2}, the diode processing consists of the following steps: oxygenation, emitter formation, backside doping, Phosphorous plateau formation, Carbon implantation and metallisation. 
The implantations were performed at HZDR Dresden \cite{HZDR}.

\begin{figure}
    \centering
    \includegraphics[width=\linewidth]{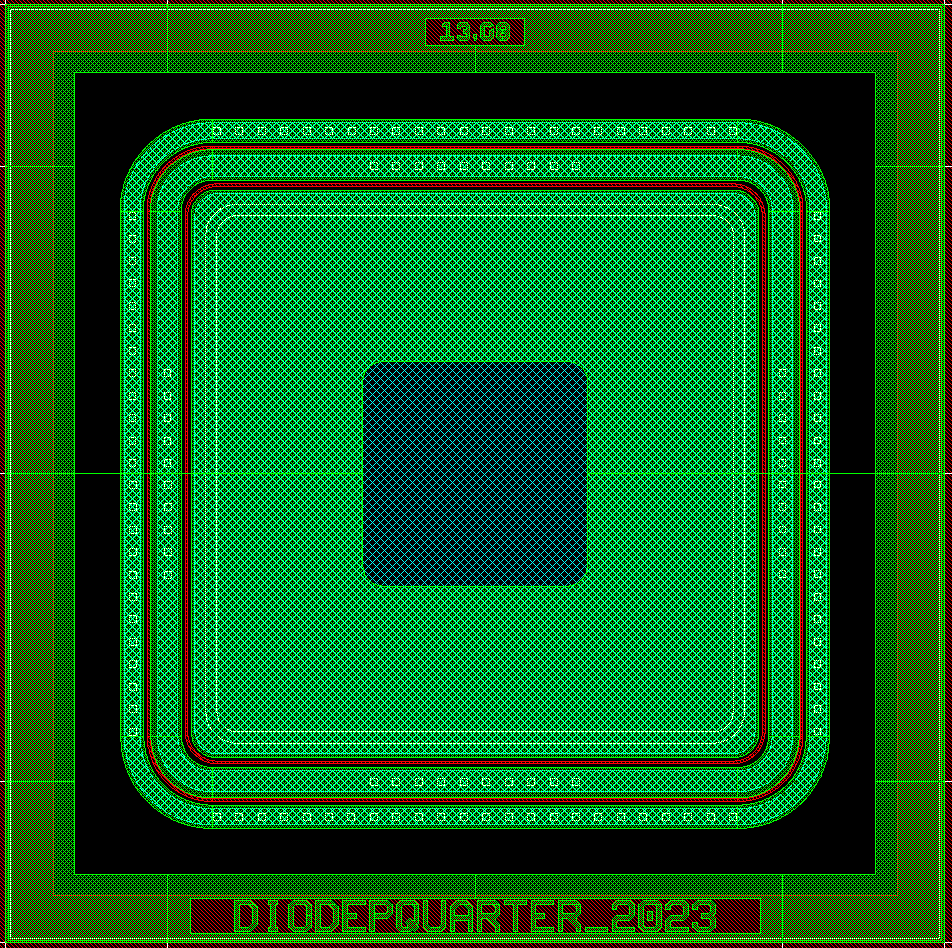}
    \caption{Layout of a small diode. The opening in the metallisation in the middle is for light injection.}
    \label{fig:layout_diode}
\end{figure}

\autoref{fig:layout_diode} shows the layout of a small diode, which only differs in its size to a large diode.

The oxygenation was performed by in-diffusion of Oxygen from an oxide layer during a prolonged annealing step (\SI{24}{\hour} at \SI{1150}{\celsius}). 
Standard Phosphorous (\SI{100}{\kilo\electronvolt}, \SI{1.5e15}{\per\centi\meter\squared}) and Boron implantations (\SI{100}{\kilo\electronvolt}, \SI{1e15}{\per\centi\meter\squared}) were made on the front and backside, respectively. 

The Phosphorous plateau was formed using four successive high-energy (\SIlist{500;1000;1250;1500}{\kilo\electronvolt}) implantations. 
The implantations were activated using an annealing step of \SI{5}{\minute} at \SI{1000}{\celsius}.

Then, a quarter of each wafer was shielded using a sacrificial quarter wafer glued on top of the wafer and high-energy (\SI{500}{\kilo\electronvolt}) Carbon implantations were made with the three aforementioned doses (\SIlist{5e12;5e13;5e14}{\per\centi\meter\squared}). 
Finally, a short annealing step was done, \SI{10}{\second} at \SI{1000}{\celsius}, using a rapid-thermal-annealing (RTA) furnace with fast infrared heater lamps.

Afterwards, the diodes were coated by an Aluminium-Silicon alloy for the metallisation. 
No passivation was applied after the metallisation.
Contact formation was performed using a low-temperature annealing step, \SI{30}{\minute} at \SI{350}{\celsius}. 
Lastly, the \IV characteristics of the diodes were measured on the wafer, then the diodes were diced.

\section{Secondary-Ion Mass Spectroscopy}

Measurements by SIMS were carried out on dedicated structures using an IMS7f-auto by Cameca~\cite{cameca} at the CiS Forschungsinstitut in Erfurt.

To avoid possible charging effects induced by surrounding non-conductive areas of the sample, a thin gold layer ($\sim$\SI{38}{\nano\meter}) was applied before loading the samples into the SIMS system. 
For the analysis of Phosphorus, Oxygen and Carbon, positively-charged Caesium ions were used as primary ions and negatively-charged secondary ions measured to ensure the best possible signal-to-noise ratio as well as low detection limits. 
The Caesium source was set at \SI{10}{\kilo\volt} and the sample at \SI{-5}{\kilo\volt}. 
High mass resolution (HMR) was applied to avoid mass interferences between different ion species with the same nominal mass: $\text{HMR} = \num{4800}$ to distinguish $^{31}$P from $^{30}$Si$^1$H and $\text{HMR} = \num{2000}$ to distinguish the reference mass $^{28}$Si from $^{16}$O$^{12}$C. 
The depth profiles were recorded until stable ion signals were reached.

The raster size was set to \num{150}$\,\times\,$\SI{150}{\microns\squared} based on the analysed structures. 
E-gating and a reduction of the energy-slit width were applied in order to reduce background and noise effects from the edge of the sputter crater.

To convert the sputtering time into a depth scale, the depth of the sputter crater was determined by a calibrated stylus profiler, Alpha-Step 500-D by KLA instruments \cite{KLA}, and a constant sputter rate assumed. 
The uncertainty of depth measurements is below \SI{10}{\percent}. 
Adequate reference samples were used to convert the measured ion intensities into atomic concentrations. 
Repeated measurements, performed under the same experimental conditions, show a maximum variation of the experimentally-determined concentration of less than \SI{2}{\percent}.

\begin{figure}[t]
    \centering
    \includegraphics[width=\linewidth]{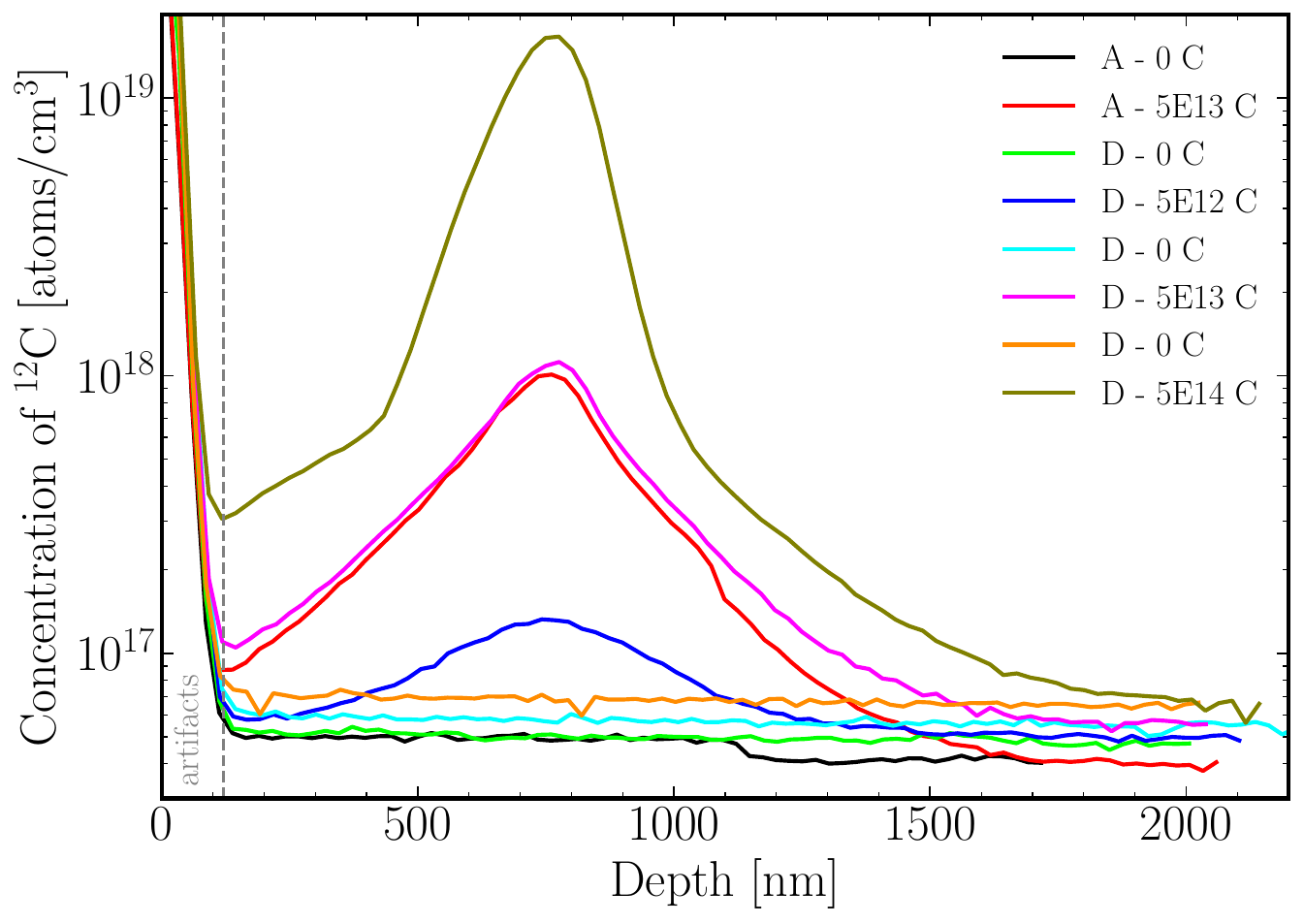}
    \caption{$^{12}$C concentration as a function of depth, obtained from SIMS measurements.}
    \label{fig:SIMS_Carbon}
\end{figure}

\begin{figure}[t]
    \centering
    \includegraphics[width=\linewidth]{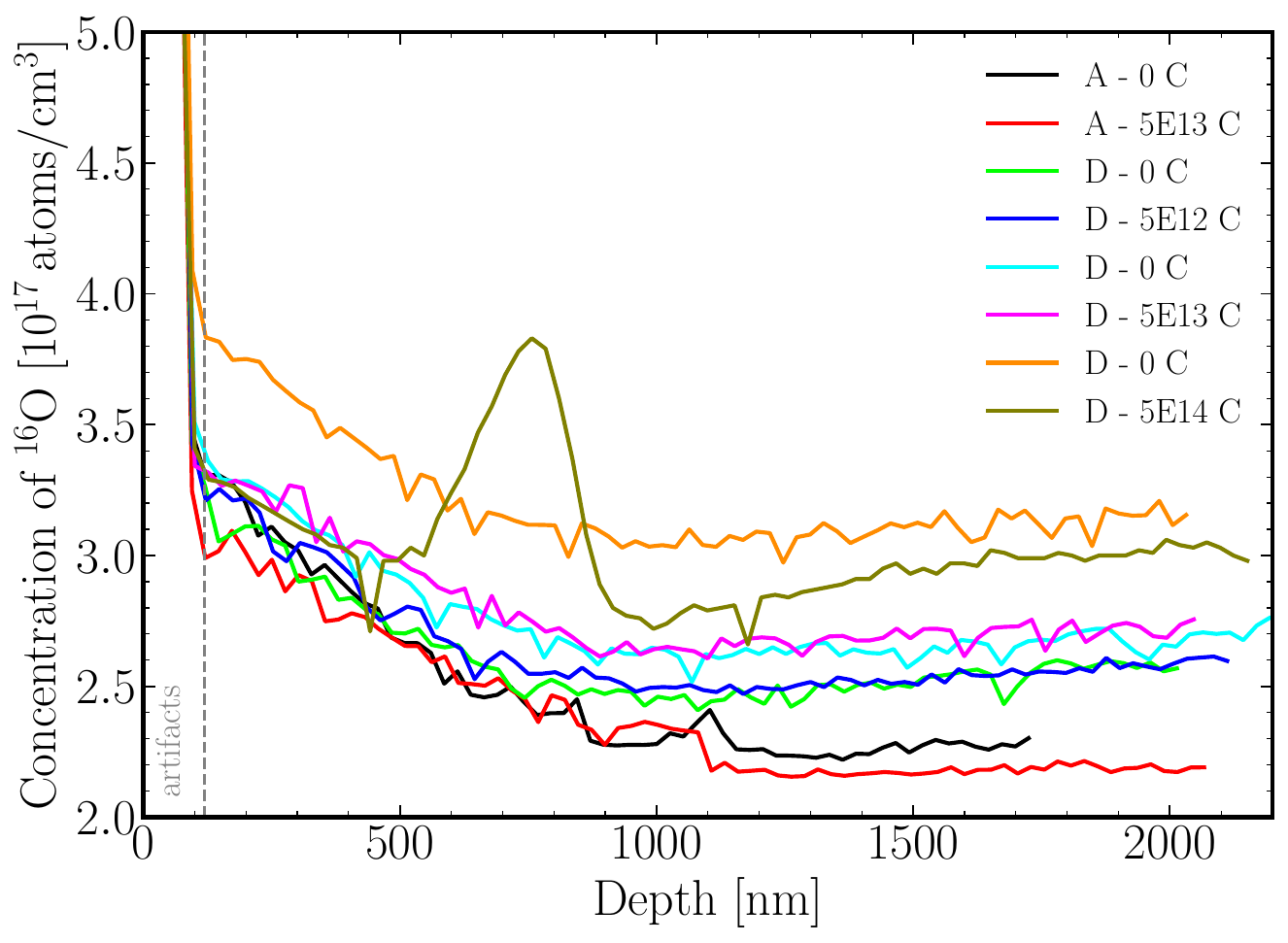}
    \caption{$^{16}$O concentration as a function of depth, obtained from SIMS measurements. It has to be noted that the measurements were done only up to a depth of around \SI{2}{\microns}, where the concentration of Oxygen in FZ and DOFZ material is still quite similar. For deeper depths, a reduction is expected for the FZ diodes, see Figure 1a in Ref.~\cite{ioana123}.}
    \label{fig:SIMS_Oxygen}
\end{figure}

As shown in \autoref{fig:SIMS_Carbon}, the Carbon concentration exhibits a peak at a depth of $\sim$\SI{750}{\nano\meter} from the surface.
The peak concentration is around \SI{1.7e19}{C\per\centi\meter\cubed} for the highest Carbon implantation dose, \SI{1e18}{C\per\centi\meter\cubed} for the middle dose and \SI{1.5e17}{C\per\centi\meter\cubed} for the lowest dose.
The measurements were performed on diodes of flavours A and D.
The measurements on the corresponding structures without any Carbon implantation show a base level of around \SI{6.e16}{C\per\centi\meter\cubed}, similar for all wafers.
\autoref{fig:SIMS_Oxygen} shows the measured Oxygen concentration.
Flavour D (DOFZ) shows a higher base-level Oxygen concentration than flavour A (FZ), as expected.

Additionally, an increase of the base Oxygen concentration (\SI{1200}{\nano\meter} and deeper) is observed for increasing Carbon concentration for the flavour D diodes.
This is also observed in the diodes without any Carbon implantation from the upper-right corner of the same wafers.
Furthermore, the measurement of the diode with the highest Carbon implantation dose shows an Oxygen peak at the same depth as the Carbon peak in \autoref{fig:SIMS_Carbon}.
No explanation has been found for either observation.

\begin{figure}[t]
    \centering
    \includegraphics[width=\linewidth]{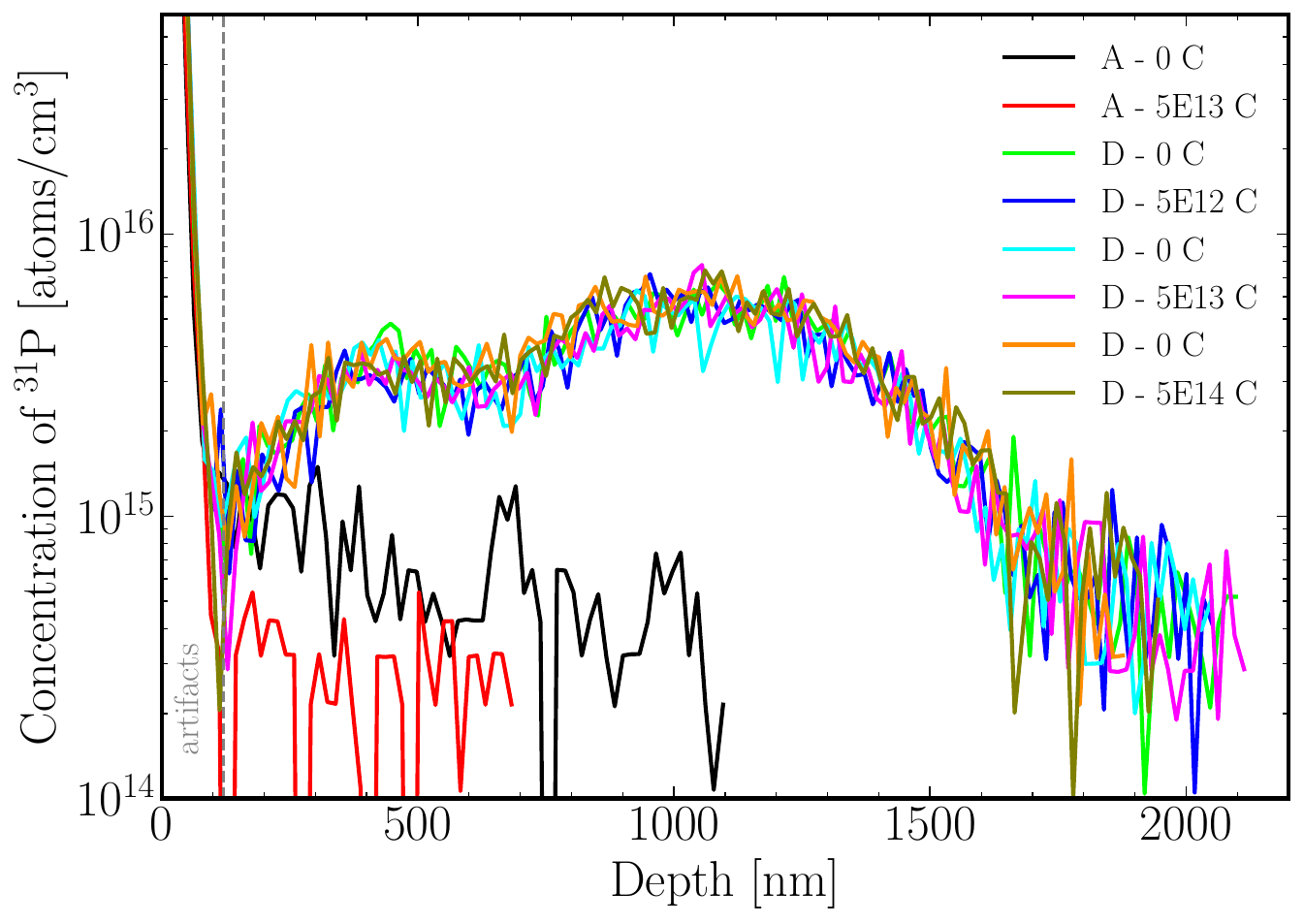}
    \caption{$^{31}$P concentration as a function of depth, obtained from SIMS measurements. The tested structures reproduce the Phosphorus plateau. The measurement sensitivity is around \SI{4e14}{\text{atoms}\per\centi\meter\cubed}.}
    \label{fig:SIMS_Phosphorus_plateau}
\end{figure}

\autoref{fig:SIMS_Phosphorus_plateau} shows the concentration of Phosphorus as a function of depth.
For flavour D diodes, the plateau is clearly visible, while for the two flavour A diodes, a much lower Phosphorus concentration is found, as expected.
For flavour D diodes, no effect of the Carbon dose is observed.

\begin{figure}[t]
    \centering
    \includegraphics[width=\linewidth]{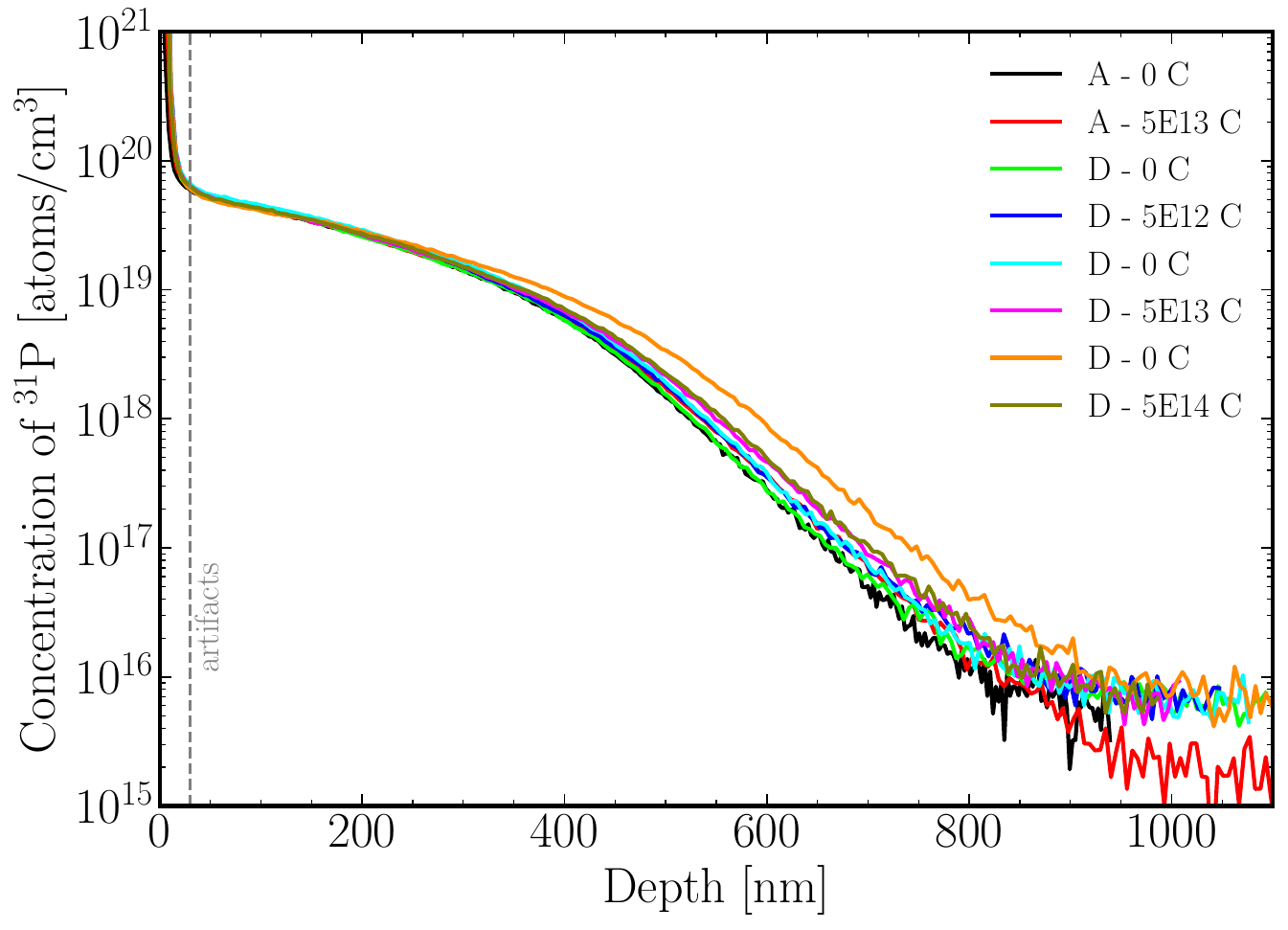}
    \caption{$^{31}$P concentration as a function of depth, obtained from SIMS measurements. The structures tested reproduce the n$^{++}$ implant. It has to be noted that these measurements were carried out on structures dedicated to SIMS measurements. The structures only have the n$^{++}$ implant, explaining the differences observed to \autoref{fig:SIMS_Phosphorus_plateau}, where structures with or without a Phosphorus plateau but no n$^{++}$ implants were measured.}
    \label{fig:SIMS_Phosphorus_doping}
\end{figure}

\autoref{fig:SIMS_Phosphorus_doping} shows the $n^{++}$ implant Phosphorus concentration.
No difference is found among the Carbon doses.
The diodes without Phosphors plateau (flavour A) have a slightly lower Phosphorus concentration at a depth of around \SI{1000}{\nano\meter}.

Except for the Oxygen concentration, the results from the measurements are as expected.
The measured Phosphorus profiles match the simulation predictions well.
A discrepancy between simulated and measured profiles is found for the Carbon concentration. 
The measured position of the Carbon concentration peak is \SI{300}{\nano\meter} below the simulated one. 
No explanation has been found for this discrepancy.

\section{\IV, \CV \& Doping Profiles}\label{sec:unirrad_IVCVdoping}
\begin{figure}[t]
    \centering
    \includegraphics[width=\linewidth]{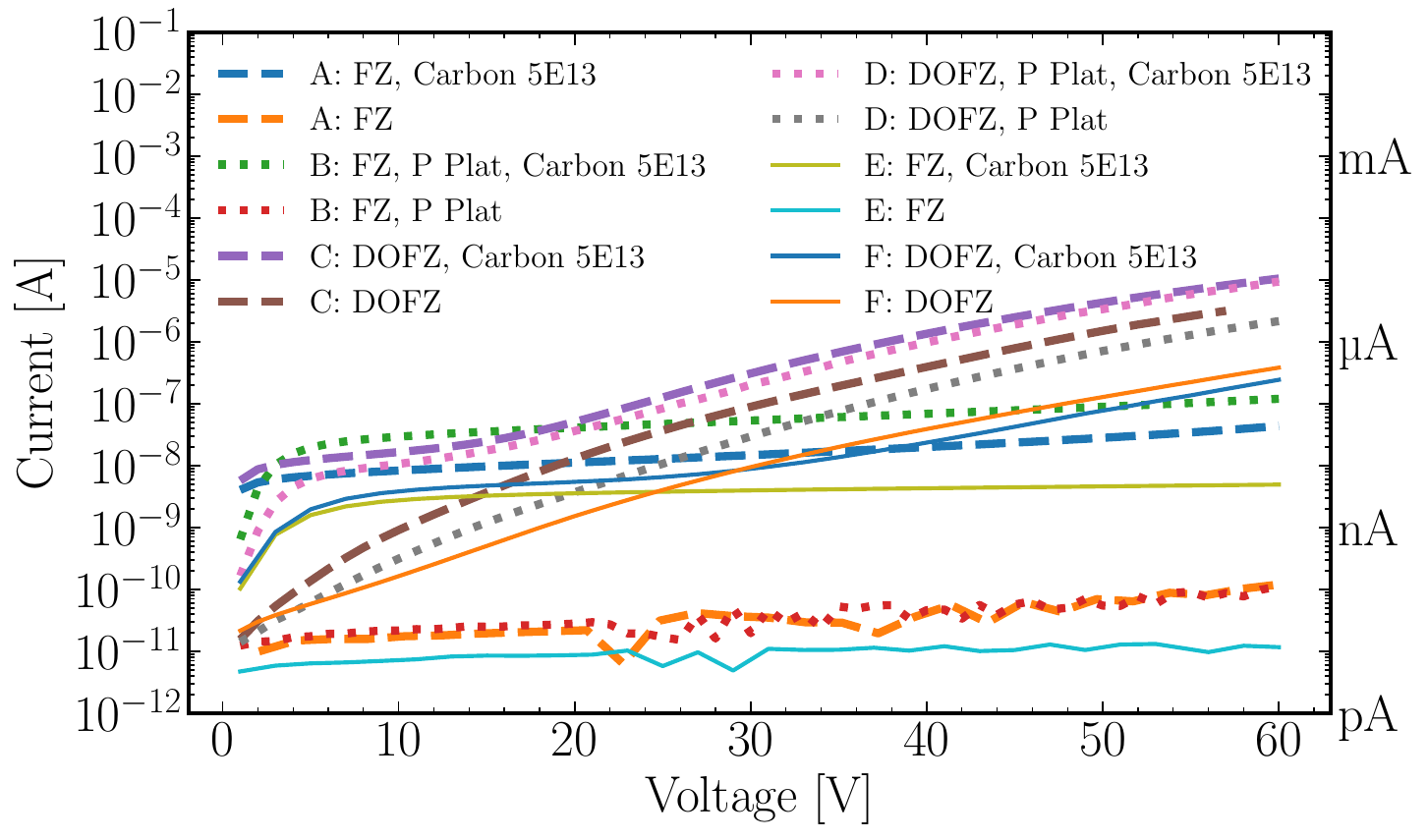}
    \caption{\IV characteristics for all flavours, with the middle Carbon dose and without Carbon.}
    \label{fig:unirrad_IV}
\end{figure}

Electrical characterisation measurements were carried out on all flavours.
The experimental setup consists of a chilled metal chuck on which the diodes are placed.
The temperature is set to \SI{20}{\celsius} for all measurements.
The high voltage is delivered to the chuck by a Keithley 2410 source meter \cite{keithley2410}.
Probe needles are placed on the contact pads of the top side of the diode and the guard ring.
The guard ring is connected to ground, while the diode's topside is connected to the low side of the high voltage line.
A Keithley 6487 pico-Ampere meter \cite{keithley6487} is placed between this return line and the sensor's topside to read out the pad current.
The current responses of the diodes are recorded while performing a bias sweep.
This setup was used for the measurements performed at the CERN SSD lab shown in \autoref{fig:unirrad_IV}.
The measurements performed at the University of Hamburg used a Keithley 6517B electrometer \cite{keithley6517B} for high voltage and pad current and a Keithley 6485 pico-Ampere meter \cite{keithley6485} for the guard ring.
The measurements shown in \autoref{fig:uniirad_IV_Hamburg_Carbon} were performed with that setup.

Examples of the \IV measurement results are shown in \autoref{fig:unirrad_IV}.
For each flavour, one diode with the middle Carbon dose (\SI{5e13}{\per\centi\meter\squared}) and one without Carbon are given as examples.
The effect of the different Carbon doses is discussed in more detail in \autoref{sec:IVCV_high_carbon}.

It has to be mentioned here that variations in the measured currents on similar samples are observed, especially for biases larger than \SI{5}{\volt}, when other mechanisms than the bulk conduction in Silicon start to contribute to the measured current. 
This is revealed by the temperature dependence of the measured current for different biases discussed in \autoref{sec:IVCV_high_carbon}. 

Comparing each FZ diode with their DOFZ counterpart in the low bias region of \autoref{fig:unirrad_IV} reveals that the DOFZ diodes all have larger leakage currents. 
Only for the carbonated diodes of flavour B and D, a larger current from the FZ diode is observed at voltages below \SI{20}{\volt}. 
However, this is only a slight increase. 

By comparing the blue and orange dashed lines in \autoref{fig:unirrad_IV}, the effect of Carbon implantation can be observed on the full range of reverse biases.
Except for their Carbon dose, these two diodes are identical and from the same wafer. The Carbon-implanted diode has a much higher leakage current than the one without Carbon. 
This increase of around three orders of magnitude is not as strongly pronounced for the DOFZ diodes since the current levels are already increased by the higher Oxygen concentration. 
In this case, the difference is only about one order of magnitude. 
Finally, differences of about one order of magnitude for the leakage currents are found between the \SI{2}{\ohmcm} and \SI{10}{\ohmcm} diodes, as expected.

\begin{figure}[t]
    \centering
    \includegraphics[width=\linewidth]{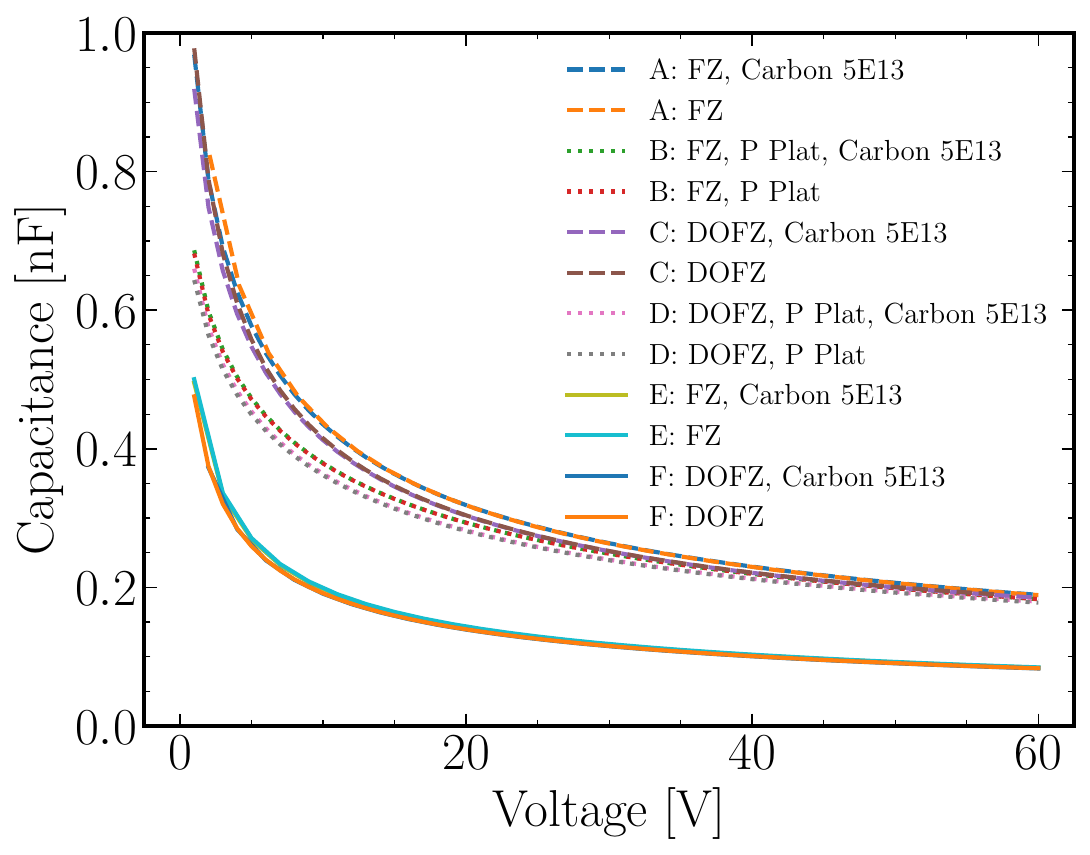}
    \caption{\CV characteristics for all flavours, with the middle Carbon dose and without Carbon.}
    \label{fig:unirrad_CV}
\end{figure}
\begin{figure}[t]
    \centering
    \includegraphics[width=\linewidth]{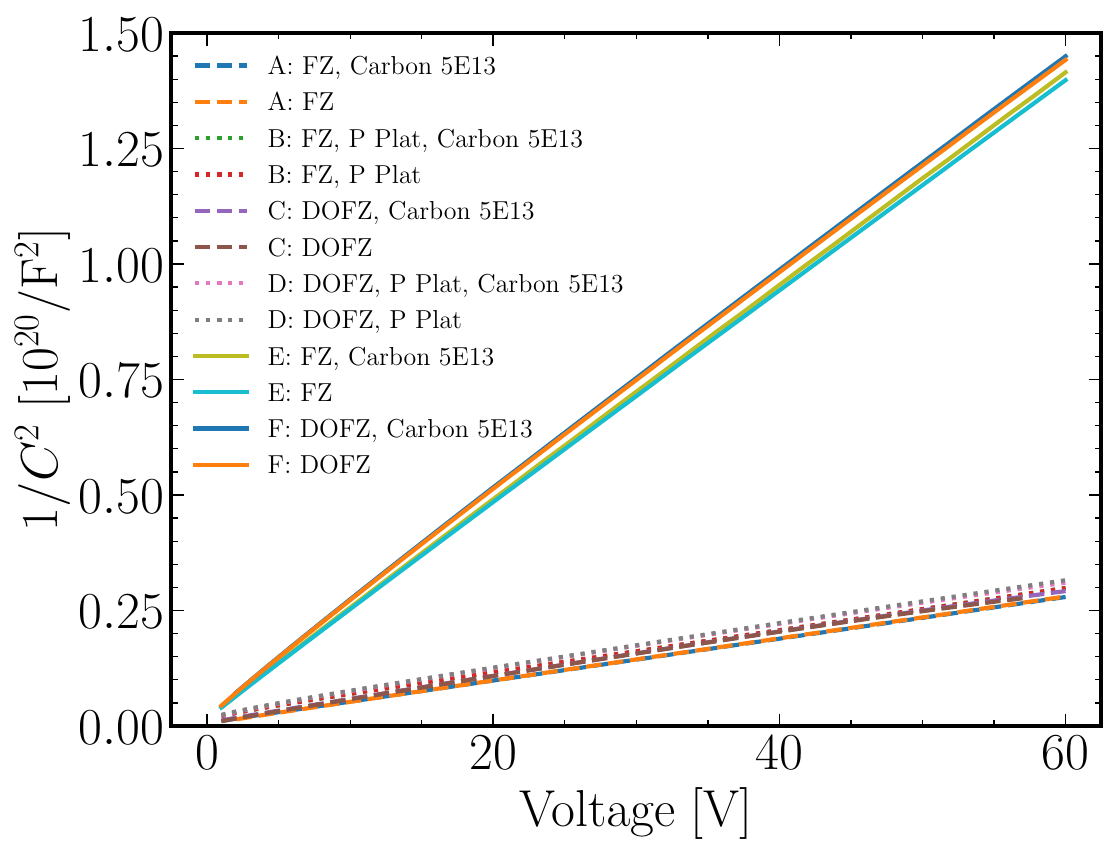}
    \caption{Inversely-squared reverse capacitance as a function of the bias voltage. Rescaling of \autoref{fig:unirrad_CV}.}
    \label{fig:unirrad_C2V}
\end{figure}
\begin{figure}[t]
    \centering
    \includegraphics[width=\linewidth]{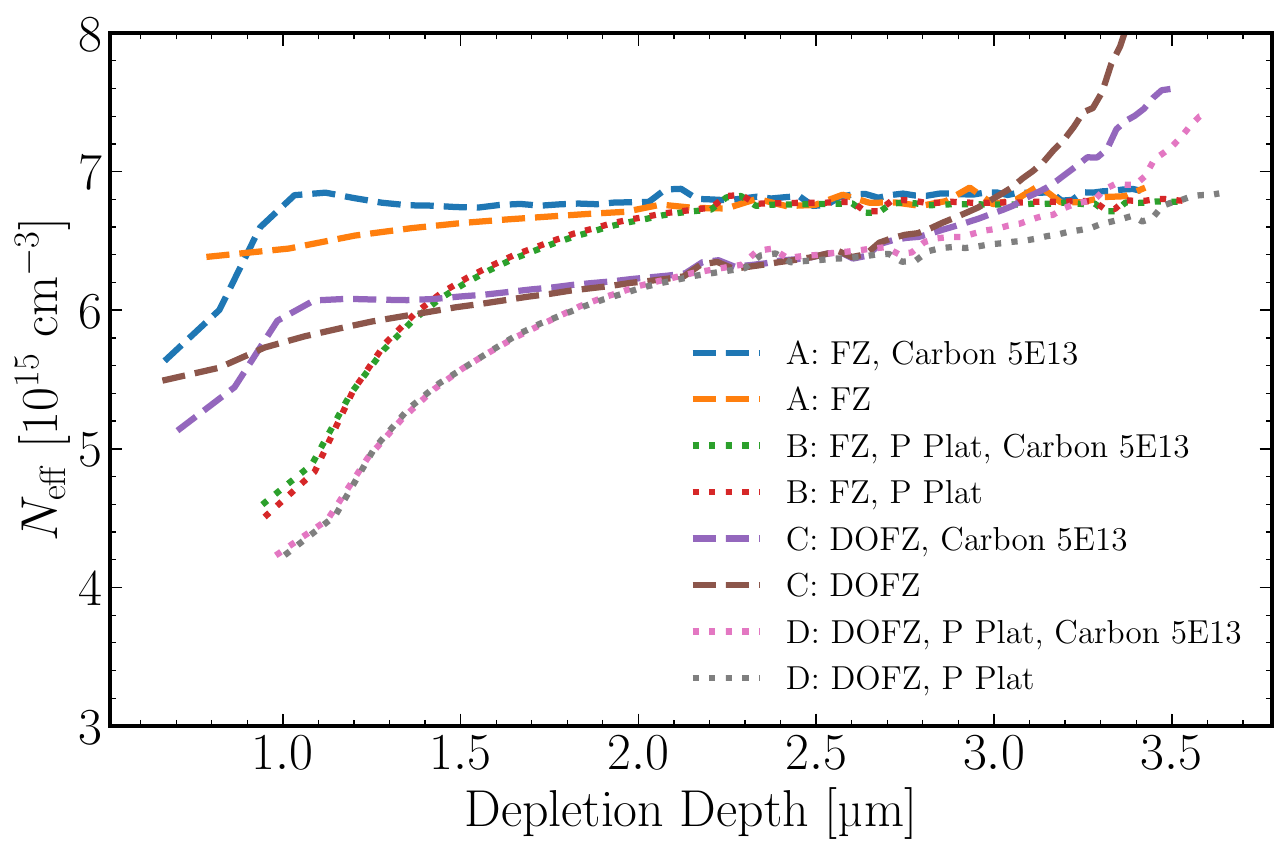}
    \caption{Effective doping profiles for all flavours of the \SI{2}{\ohmcm} diodes, with the middle Carbon dose and without Carbon, obtained from the \CV measurements.}
    \label{fig:unirrad_doping_2Ohmcm}
\end{figure}
\begin{figure}[t]
    \centering
    \includegraphics[width=\linewidth]{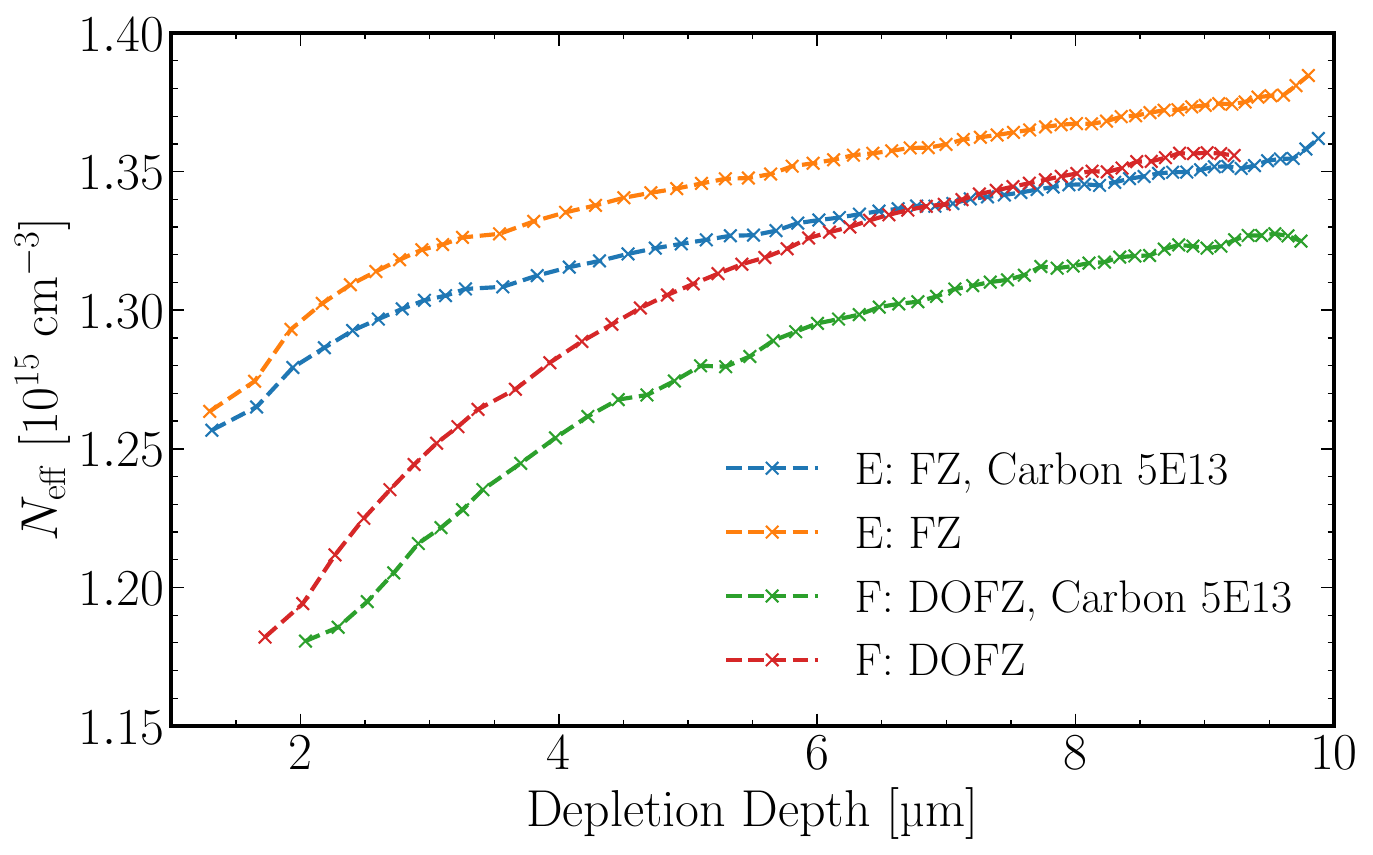}
    \caption{Effective doping profiles for all flavours of the \SI{10}{\ohmcm} diodes, with the middle Carbon dose and without Carbon, obtained from \CV measurements.}
    \label{fig:unirrad_doping_10Ohmcm}
\end{figure}

To measure the \CV characteristics and the doping profiles, the setup at CERN used an Agilent E4980A LCR meter \cite{agilent} and a Keithley 2410 source meter.
The measurements shown in \figuresref{fig:unirrad_CV}, \ref{fig:unirrad_C2V}, \ref{fig:unirrad_doping_2Ohmcm} and \ref{fig:unirrad_doping_10Ohmcm} were performed using that setup.
The measurements taken at the University of Hamburg, shown in \autoref{fig:unirrad_doping_Hamburg_Carbon}, were performed using an Agilent E4980A LCR meter in combination with a Keithley 6517B electrometer.

The capacitance as a function of the reverse bias is shown in \autoref{fig:unirrad_CV}.
The measurements were taken at a frequency of \SI{10}{\kilo\hertz} and an AC amplitude of \SI{500}{\milli\volt}.
The diodes with \SI{10}{\ohmcm} resistivity have noticeably lower capacitance than the ones with \SI{2}{\ohmcm} resistivity.
This is also observed in \autoref{fig:unirrad_C2V}, where the inverse square of the capacitance as a function of the bias voltage is shown.
Assuming a parallel plate capacitor equivalent for a diode, a linear increase is expected until full depletion is reached.
The measurements match this expectation.

The effective doping concentration \Neff is calculated from the slope of a linear fit on the increase $1/C^2(V)$ using 
\begin{align}
    N_\text{eff}=-\dfrac{2}{\varepsilon_0\varepsilon_rA^2q_0\dfrac{\text{d}\left(1/C^2\right)}{\text{d}V}},\label{eq:Neff1}
\end{align}
where $q_0$ is the elementary charge, $A$ the surface area of the diode encased by the guard ring and $\varepsilon_0$ and $\varepsilon_r$ the vacuum and relative permittivity of Silicon, respectively.

The capacitance is measured at up to \SI{60}{\volt}, even though full depletion is not reached.
Knowing \Neff, the full depletion voltage \Vdep can be estimated using
\begin{align}
    V_\text{dep}=\dfrac{q\left|N_\text{eff}\right|d^2}{2\varepsilon_0\varepsilon_r},
\end{align}
where $d$ is the physical thickness of the diode.
A mean value of around \SI{318}{\kilo\volt} is found for the \SI{2}{\ohmcm} diodes and \SI{64}{\kilo\volt} for the \SI{10}{\ohmcm} ones.
These biases are much higher than the breakdown voltages of the diodes.
Therefore, it is impossible to fully deplete these diodes without destroying them.
This does not prevent capacitance-based DLTS measurements.

As shown in both figures, the diodes with the Phosphorus plateau co-doping have a lower capacitance compared to their counterparts without Phosphorus.
This is expected, since the compensation doping reduces \Neff, which is the difference between the donor and the acceptor concentration.
This can also be seen in \autoref{tab:unirrad_Neff}, where the \Neff extracted from the slope of the increase in \autoref{fig:unirrad_C2V} are listed for all diodes.

\begin{table}[]
\centering
\resizebox{\linewidth}{!}{%
\begin{tabular}{lc}
\hline
Flavour & \Neff [\SI{e15}{\per\centi\meter\cubed}] \\ \hline
A: FZ, Carbon 5E13 & 6,80 \\
A: FZ & 6,75 \\
B: FZ, P Plat, Carbon 5E13 & 6,70 \\
B: FZ, P Plat & 6,72 \\
C: DOFZ, Carbon 5E13 & 6,46 \\
C: DOFZ & 6,47 \\
D: DOFZ, P Plat, Carbon 5E13 & 6,43 \\
D: DOFZ, P Plat & 6,36 \\\hline
E: FZ, Carbon 5E13 & 1,33 \\
E: FZ & 1,34 \\
F: DOFZ, Carbon 5E13 & 1,30 \\
F: DOFZ & 1,31 \\ \hline
\end{tabular}%
}
\caption{\Neff calculated from the slope of \CV measurements in \autoref{fig:unirrad_C2V}. Flavours A through D have a nominal resistivity of \SI{2}{\ohmcm}, flavours E and F \SI{10}{\ohmcm}.}
\label{tab:unirrad_Neff}
\end{table}

The effective doping concentration as a function of the depleted depth is shown in \figuresref{fig:unirrad_doping_2Ohmcm} and \ref{fig:unirrad_doping_10Ohmcm} for the \SIlist{2;10}{\ohmcm} diodes, respectively.
This is obtained from \CV measurements and
\begin{align}
    N_\text{eff}(x)=-\dfrac{2}{q_0\varepsilon_0\varepsilon_rA^2}\left[\dfrac{\text{d}}{\text{d}V}\left(\dfrac{1}{C^2(x)}\right)\right]^{-1},\label{eq:Neff2}
\end{align}
where $V$ is the voltage and $x$ the depleted depth, calculated from the \CV measurements as
\begin{align}
    x(V)=\dfrac{\varepsilon_0\varepsilon_rA}{C(V)}.
\end{align}
\autoref{eq:Neff2} is essentially \autoref{eq:Neff1}, however, the slope is evaluated at each point $x$ and not extracted from a linear fit, as is done for in \autoref{eq:Neff1} and \autoref{fig:unirrad_C2V}.
In \autoref{fig:unirrad_doping_2Ohmcm} a clear effect of the Phosphorus plateau implantation is visible.

Comparing a diode with Phosphorus to one without Phosphorus (orange dashed versus green dotted line) a reduced effective doping concentration up to a depth of around \SI{2}{\microns} is observed.
This is consistent with the SIMS measurements in \autoref{fig:SIMS_Phosphorus_plateau}.

An effect of the Carbon implantation is visible by comparing the blue with the orange dashed line.
It appears, that the Carbon implantation reduces the effective doping concentration, up to a depth of around \SI{1}{\micron}, which is consistent with the Carbon implantation depth observed by the SIMS measurements in \autoref{fig:SIMS_Carbon}.
This is expected due to the partial activation of the acceptor induced by the Carbon \cite{marcoferrero}.
However, this effect is only observed in diodes without the Phosphorus implant.
Comparing the green and red dotted lines in \autoref{fig:unirrad_doping_2Ohmcm}, no significant difference is observed.
Lastly, FZ diodes have a larger effective doping concentration than their DOFZ counterparts.

\subsection{Comparison of Carbon doses}\label{sec:IVCV_high_carbon}
\begin{figure}[t]
    \centering
    \includegraphics[width=\linewidth]{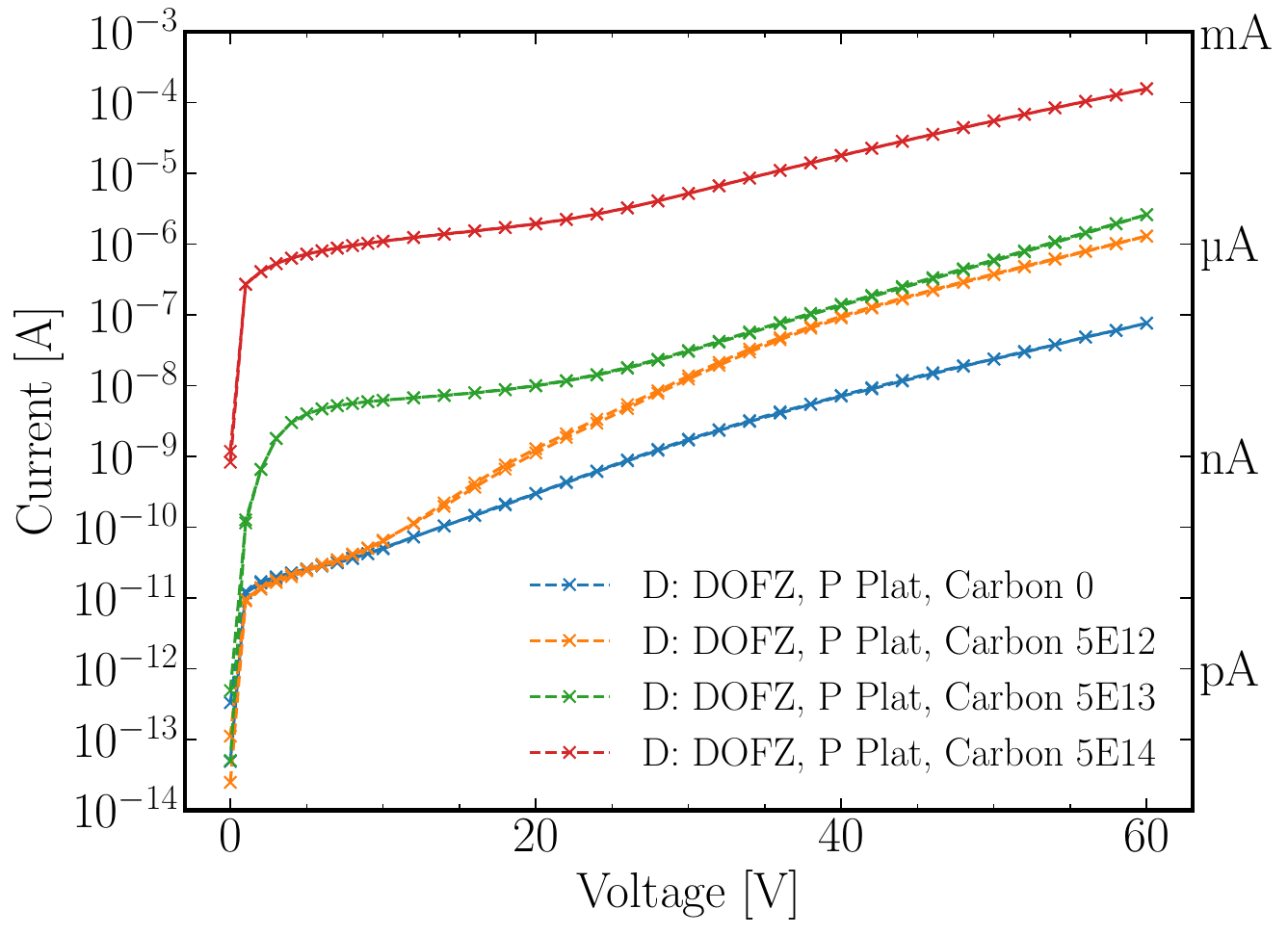}
    \caption{Comparison of the leakage currents as a function of applied reverse bias for diodes of flavour D with different Carbon implantation doses.}
    \label{fig:uniirad_IV_Hamburg_Carbon}
\end{figure}

A comparison of the effect of the Carbon implantation dose on the leakage currents is shown in \autoref{fig:uniirad_IV_Hamburg_Carbon} for diodes of flavour D (\SI{2}{\ohmcm}, DOFZ and with Phosphorus plateau).
The currents increase with increasing Carbon doses.
However, the currents originating from the diode with the lowest Carbon dose are compatible with the currents from the middle dose for reverse biases above \SI{40}{\volt}.
Increased leakage currents with Carbon implantation has been observed in the past \cite{carbonated_more_current1,carbonated_more_current2,carbonated_more_current3}.
The currents are measured both while increasing and decreasing the bias voltage. 
Both scans are shown in \autoref{fig:uniirad_IV_Hamburg_Carbon}.
No hysteresis is observed.

On the other hand, the shape of the \IV characteristics differs from that of ideal diodes, suggesting that not only the bulk conduction in Silicon contributes to the measured leakage current. 
The reverse current versus temperature was measured to extract the current activation energies for different biases. 
This way it was possible to distinguish two different conduction mechanisms. 
The first one corresponds to the bulk conduction of Silicon, with an activation energy of around \SI{1.25}{\electronvolt}, clearly measurable at low reverse biases (e.g. \SIrange{5}{10}{\volt}).
The second one, with an activation energy varying between \SIlist{0.06;0.6}{\electronvolt} depending on the sample and temperature, starts to contribute to the conduction at increased biases. 
\begin{figure}
    \centering
    \includegraphics[width=\linewidth]{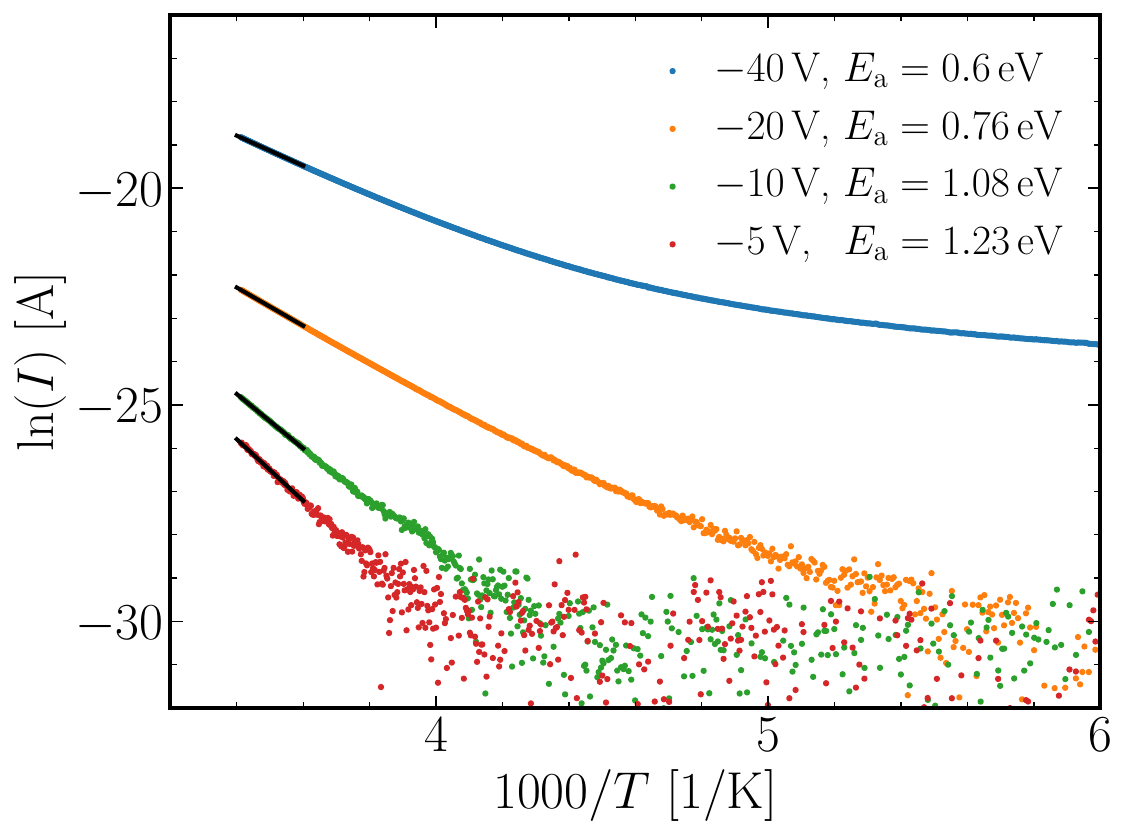}
    \caption{Example of the temperature dependence of the leakage current for different applied reverse biases. The corresponding activation energies extracted from linear fits to the data points (shown in black) are stated in the legend.}
    \label{fig:uniirad_IV_activation_bucharest}
\end{figure}
An example revealing these conduction mechanisms is given in \autoref{fig:uniirad_IV_activation_bucharest} for a flavour D diode with a Carbon implantation of \SI{5e12}{\per\centi\meter\squared}. 
Thus, in most of the samples a large current contribution from surface effects which are activated for biases larger than \SI{5}{\volt} distort the \IV characteristic from an ideal shape. 
This conduction mechanism at larger biases is associated to surface related defects located most likely between the pad and the guard rings. 
The measurements show that the generated surface current may vary from sample to sample, irrespective the amount of Oxygen or implanted Carbon and Phosphorous.

\begin{figure}
    \centering
    \includegraphics[width=\linewidth]{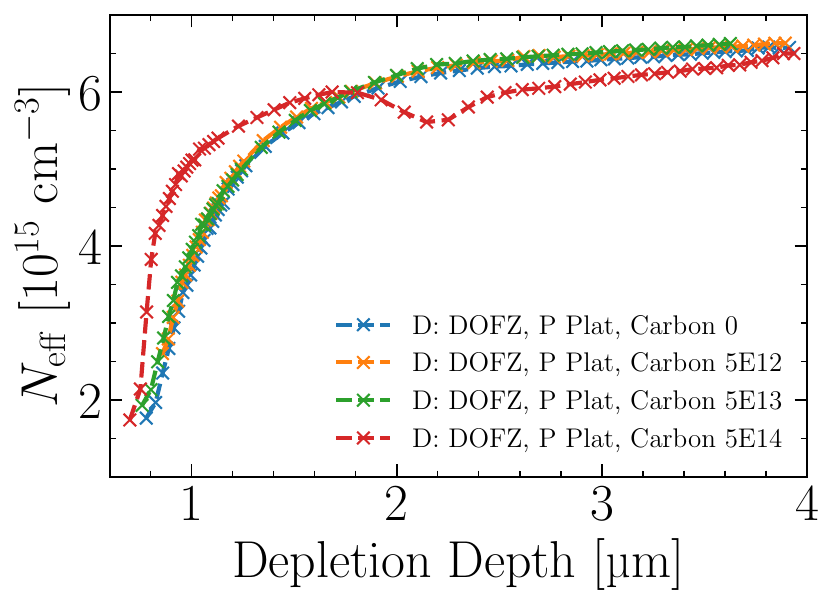}
    \caption{Comparison of the doping profiles as a function of depth, obtained from \CV measurements for different Carbon implantation doses for flavour D diodes.}
    \label{fig:unirrad_doping_Hamburg_Carbon}
\end{figure}

The effect of the Carbon dose on the doping profile, exemplarily for flavour D diodes, is shown in \autoref{fig:unirrad_doping_Hamburg_Carbon}.
All doping profiles are compatible with each other, except for the one of the diode with the highest carbon dose.
Similar behaviour was observed for the other flavours as well.
No explanation for this has been found. 
It is noted that in this case even the depth dependence of the doping profile shown in \autoref{fig:unirrad_doping_Hamburg_Carbon} might be affected by errors as the analysis based on \autoref{eq:Neff1} assumes a monotone extension of the depleted volume from the surface into the bulk with rising voltage. 
If the Carbon is locally compensating some of the space charge, this might no longer be true and lead to artifacts in the doping depth profile.

To conclude, the measurements show that the diodes, except for the ones with the highest Carbon dose, behave as expected.
Expected depletion behaviour and no early breakdowns are observed.
The measured doping profiles are in line with expectations and the SIMS measurements.

\section{Deep-Level Transient Spectroscopy}
DLTS is a technique in which capacitance transients caused by the de-trapping of charges from defect states are measured.
The recorded capacitance transients at each measuring temperature are analysed by using several correlator functions (discrete Fourier analysis) \cite{dlts1, dlts2}.   
This yields time constants which follow an Arrhenius relation, which in turn is used to extract the energy level and capture cross-section of defects.
Furthermore, the defect concentrations are extracted.
More details about this method can be found in Ref.~\cite{Moll}.

To achieve comparability between DLTS measurements performed at the various participating institutes, a standard set of measurement parameters was created used for all measurements.
For standard measurements, while the majority-and-minority charge carrier injection was measured with a filling pulse of \SI{-2}{\volt} forward bias for \SI{100}{\milli\second}, the majority-only charge carrier injection was measured with \SI{0.6}{\volt} reverse bias.
Three time windows \Tw are always measured: \SIlist{19.2;192;1920}{\milli\second}.
The measurements are taken at a reverse bias of \SI{10}{\volt}, with a heating rate of \SI{0.1}{\kelvin\per\second} and a minimum averaging measurement time of \SI{1}{\second} for each transient.
The measured temperature range is typically from \SIrange{20}{300}{\kelvin}.

\begin{figure}[t]
    \centering
    \includegraphics[width=\linewidth]{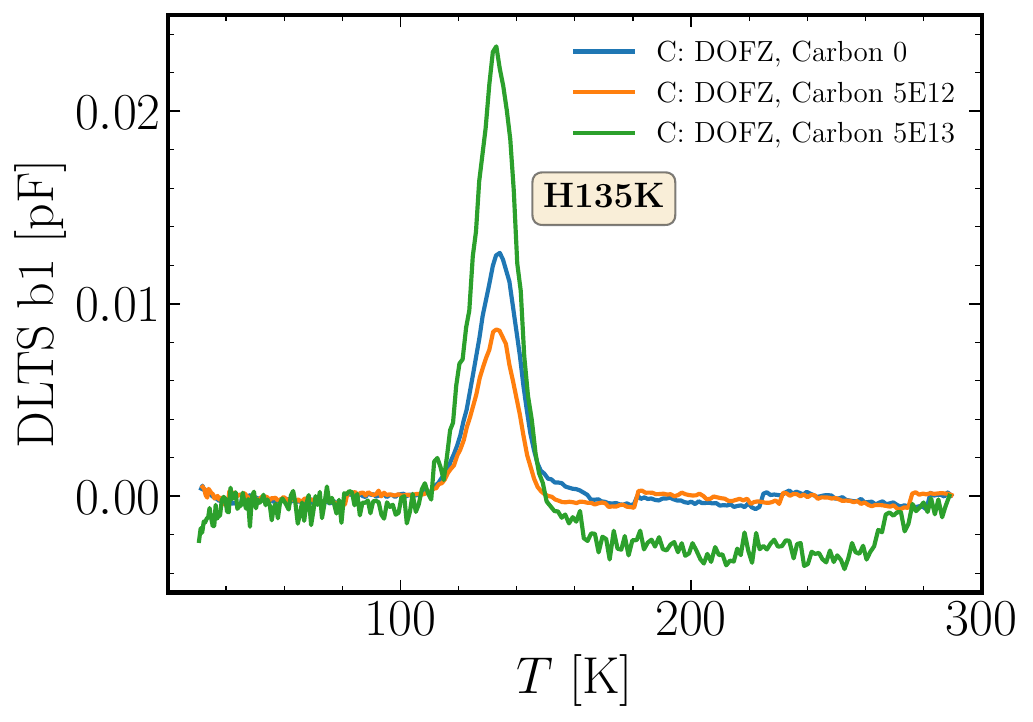}
    \caption{DLTS measurement spectra obtained for flavour C diodes with no Carbon and low and middle Carbon doses.}
    \label{fig:unirrad_DLTS_1}
\end{figure}

DLTS measurements were carried out on all flavours and all Carbon doses.
For all measurements, the spectra look nearly identical.
All spectra feature a hole trap at around \SI{135}{\kelvin}, labelled \textit{H135K}.
In the temperature range \SIrange{20}{300}{\kelvin}, no other peaks are observed.
\autoref{fig:unirrad_DLTS_1} shows, as an example, the spectra corresponding to the sine based correlator b1 obtained from a few \SI{2}{\ohmcm} samples and a time window \Tw of \SI{192}{\milli\second}.
The amplitude, and therefore the defect concentration \NT, depends on the Carbon concentration, but not in a systematic way.
The amplitude for the diode without Carbon is between that for the diodes with the lowest and middle Carbon doses.
The diode flavour also affects the amplitude, also in a non-systematic way.

\textit{H135K} has not been identified with any known defect complex.
The extracted energy level \EA and capture cross-section $\sigma$ are \SI{0.27(1)}{\electronvolt} and \SI{6(3)e-15}{\centi\meter\squared}.
With these trapping parameters the \textit{H135K} defect has practically no influence on the measured leakage currents.

\begin{figure}[t]
    \centering
    \includegraphics[width=\linewidth]{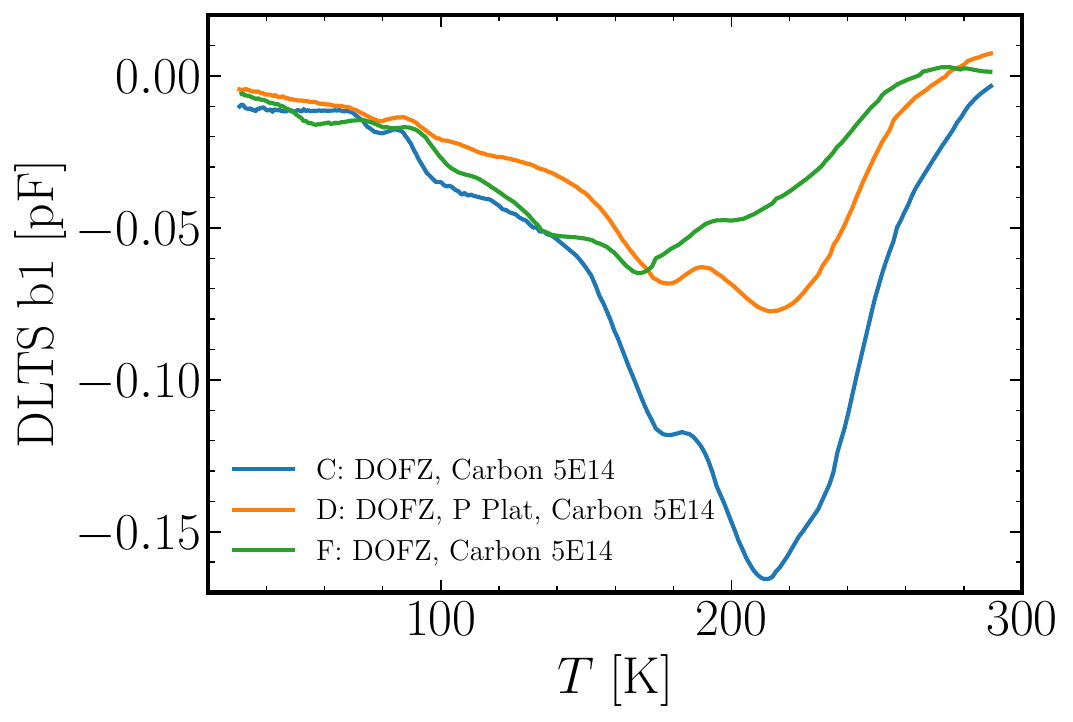}
    \caption{DLTS measurement spectra obtained from diodes of flavour C, D and F, all with the highest Carbon implantation dose. }
    \label{fig:unirrad_DLTS_2}
\end{figure}

For the highest Carbon dose the spectra are significantly different.
As shown in \autoref{fig:unirrad_DLTS_2}, broad negative peaks appear.
This is true for all flavours.
The reverse capacitance as a function of temperature, recorded during the measurement, shows no strong change over the measured temperature range, therefore also not explaining the broad shapes.
No explanation has been found for these results.
From the SIMS, DLTS, \IV and \CV measurements, it is clear that the diodes with the highest Carbon dose have characteristics that are not well understood.

\section{Conclusion}
The problem of gain-layer degradation in LGADs is of vital importance for future HEP experiments.
For detectors such as the High-Granularity Timing Detector (HGTD) of the ATLAS experiment, which uses carbonated LGADs, two replacements are planned during the operation in the High-Luminosity LHC (HL-LHC).
An increased radiation hardness of the gain layer would be beneficial for such experiments, reducing the costs.

Concepts such as carbonated LGADs have been proven to increase the radiation hardness of the gain-layer.
However, the beneficial role of the addition of Carbon is only superficially understood, and there is no complete understanding of the defect kinetics leading to the improved radiation hardness.
Further experimental studies on defect formation are paramount to gain this knowledge. 
This is the aim of the Gain-Layer Project.

\num{19050} diodes with various doping schemes have been produced.
The doping concentration of these diodes' bulks is very close to that of typical LGAD gain-layers.
These diodes therefore, enable a direct study of gain-layer-like structures with defect spectroscopy methods such as DLTS and TSC.
The various doping schemes allow to study the effect of various Carbon implantation doses, resistivities (Boron concentrations), Phosphorus co-doping and Oxygen concentrations.

This article introduces the project, describes the diode processing in detail and reports pre-irradiation results from SIMS, \IV, \CV and DLTS measurements.
DLTS measurements revealed one hole trap at around \SI{135}{\kelvin} of unknown structure. 
The concentration is low and will therefore not impact the upcoming irradiation experiments. 

The diodes with the highest Carbon implantation dose of \mbox{\SI{5e14}{\per\centi\meter\squared}} gave unexpected SIMS results for the Oxygen concentration, showing a peak at the same depth as the Carbon implantation.
The doping profiles also deviate from those of the diodes with all other Carbon implantation doses, as well as the diodes with no Carbon implantation.
DLTS measurements show minority carrier traps for majority-carrier-only-injection measurements.
Furthermore, extremely broad peak are seen, and this could not be explained by large changes in the reverse capacitance with temperature.
No explanation for these observations has been found

All other diode flavours work as expected.
In the future, various irradiation campaigns will be performed to study the acceptor removal effect as a function of the doping schemes and irradiation fluences.
Proton and neutron irradiation campaigns are planned, where fluence effects can be observed and the damage compared between irradiation type and doping scheme.
Furthermore, extreme-fluence irradiations are planned (\SI{e17}{\neqpcm} and higher).
Irradiated devices will be investigated with \IV, \CV, DLTS, TSC, TSCap, PL, Hall and FTIR.

\FloatBarrier
\appendix


\section*{Acknowledgements}
The work was performed partly in the framework of the CERN RD50 and DRD3 collaborations.

Niels Sorgenfrei acknowledges that his work has been sponsored by the Wolfgang Gentner Programme of the German Federal Ministry of Research, Technology and Space (grant no. 13E18CHA).

Ioana Pintilie acknowledges the funding received through the IFA-CERN-RO 07/2024 project.

Tomas Ceponis acknowledges the funding received through the agreement of Vilnius University with the Lithuanian Research Council No. VS-13.

\section*{Declaration of competing interest}
The authors declare that they have no known competing financial interests or personal relationships that could have influenced the work reported in this paper.

\section*{Data availability}
The data will be made available on request.

\section{Additional Table}
\begin{table*}[]
\centering
\resizebox{\textwidth}{!}{%
\begin{tabular}{ccccccccc}
\hline
Flavour & Wafer & Run & Thickness & Type & Resistivity & DOFZ & P Plateau & C Dose \\
 & Number & Number & [\si{\microns}] &  & [\si{\ohmcm}] &  &  & [\si{\per\centi\meter\squared}] \\ \hline
A & 1  & \num{440598} & 250 & FZ &  2 & no  & no  & \num{5E+12} \\
A & 2  & \num{440598} & 250 & FZ &  2 & no  & no  & \num{5E+13} \\
A & 3  & \num{440598} & 250 & FZ &  2 & no  & no  & \num{5E+13} \\
A & 4  & \num{440598} & 250 & FZ &  2 & no  & no  & \num{5E+13} \\
A & 5  & \num{440598} & 250 & FZ &  2 & no  & no  & \num{5E+14} \\
B & 6  & \num{440598} & 250 & FZ &  2 & no  & yes & \num{5E+12} \\
B & 7  & \num{440598} & 250 & FZ &  2 & no  & yes & \num{5E+13} \\
B & 8  & \num{440598} & 250 & FZ &  2 & no  & yes & \num{5E+13} \\
B & 9  & \num{440598} & 250 & FZ &  2 & no  & yes & \num{5E+14} \\
C & 10 & \num{440598} & 250 & FZ &  2 & yes & no  & \num{5E+12} \\
C & 11 & \num{440598} & 250 & FZ &  2 & yes & no  & \num{5E+13} \\
C & 12 & \num{440598} & 250 & FZ &  2 & yes & no  & \num{5E+13} \\
C & 13 & \num{440598} & 250 & FZ &  2 & yes & no  & \num{5E+14} \\
D & 14 & \num{440598} & 250 & FZ &  2 & yes & yes & \num{5E+12} \\
D & 15 & \num{440598} & 250 & FZ &  2 & yes & yes & \num{5E+13} \\
D & 16 & \num{440598} & 250 & FZ &  2 & yes & yes & \num{5E+13} \\
D & 17 & \num{440598} & 250 & FZ &  2 & yes & yes & \num{5E+14} \\ \hline
E & 1  & \num{440603} & 525 & FZ & 10 & no  & no  & \num{5E+12} \\
E & 2  & \num{440603} & 525 & FZ & 10 & no  & no  & \num{5E+13} \\
E & 3  & \num{440603} & 525 & FZ & 10 & no  & no  & \num{5E+13} \\
E & 4  & \num{440603} & 525 & FZ & 10 & no  & no  & \num{5E+14} \\
F & 5  & \num{440603} & 525 & FZ & 10 & yes & no  & \num{5E+12} \\
F & 6  & \num{440603} & 525 & FZ & 10 & yes & no  & \num{5E+13} \\
F & 7  & \num{440603} & 525 & FZ & 10 & yes & no  & \num{5E+13} \\
F & 8  & \num{440603} & 525 & FZ & 10 & yes & no  & \num{5E+14} \\ \hline
\end{tabular}%
}
\caption{Overview of all wafers and flavours from the Gain-Layer Project. One quarter for each wafer was left without Carbon implantation.}
\label{tab:glpd}
\end{table*}

\FloatBarrier
\bibliographystyle{elsarticle-num} 
\bibliography{literature}

\end{document}